\documentclass[12pt,a4paper]{article}

\usepackage{amsfonts}
\usepackage{amsmath}
\usepackage{amssymb}
\usepackage{latexsym}

\usepackage[dvips]{graphicx}

\numberwithin{equation}{section}

\newcommand{\lanln}[1]{$\langle$\texttt{arXiv:#1}$\rangle$}

\newcommand{\BbbR}{\mathbb{R}}

\title{Transition rate of the Unruh-DeWitt detector\\
in curved spacetime}
\author{Jorma Louko\thanks{jorma.louko@nottingham.ac.uk}
\ and
Alejandro Satz\thanks{pmxas3@nottingham.ac.uk}
\\
\noalign{\vspace{3ex}}
\small{\it School of Mathematical Sciences,
University of Nottingham,}\\
\small{\it Nottingham NG7 2RD, UK}
\\
\noalign{\vspace{3ex}}
\small{Revised November 2007}
\\
\noalign{\vspace{1ex}}
\small{Published in Class.\ Quantum Grav.\ \textbf{25} (2008) 055012}
}

\date{}

\begin{document}

\maketitle

\begin{abstract}
We examine the Unruh-DeWitt particle detector coupled to a scalar
field in an arbitrary Hadamard state in four-dimensional curved
spacetime. Using smooth switching functions to turn on and off the
interaction, we obtain a regulator-free integral formula for the total
excitation probability, and we show that an instantaneous transition
rate can be recovered in a suitable limit. Previous results in
Minkowski space are recovered as a special case. As applications, we
consider an inertial detector in the Rindler vacuum and a detector at
rest in a static Newtonian gravitational field. Gravitational
corrections to decay rates in atomic physics laboratory experiments on
the surface of the Earth are estimated to be suppressed by 42 orders
of magnitude.
\end{abstract}

\section{Introduction}

The Unruh-DeWitt model for a particle detector \cite{unruh,DeWitt} is
an important tool for probing the physics of quantum fields wherever
noninertial observers or curved backgrounds are present. In such cases
there is often no distinguished notion of a ``particle,'' analogous to
the plane-wave modes in Minkowski space, but an operational meaning
can be attached to the concept by analysing the transitions induced
among the energy levels of a detector coupled to the field. Upwards or
downwards transitions can then be interpreted as due to absorption or
emission of field quanta, or particles. The best-known applications of
this procedure are those for which the spectrum of transitions is
thermal, which is the case for uniformly accelerated detectors in
Minkowski space~\cite{unruh}, inertial detectors in de Sitter space
\cite{gibb-haw:dS}, and detectors at rest in the exterior
Schwarzschild black hole spacetime~\cite{hawking}. 

In first-order perturbation theory, the transition probability of the
Unruh-DeWitt detector is proportional to a quantity known as the
response function, which involves integrating the Wightman distribution of
the quantum field over the worldline of the detector. When the quantum
state of the field is sufficiently regular and the detector is
switched on and off smoothly, the response function is well
defined~\cite{Junker}, and the physical interpretation is that the
response function is then proportional to the probability of a
transition to have occurred by a time at which all interaction has
already ceased.  If, however, one wishes to address the probability of
a transition to have occurred by a time at which the interaction is
still ongoing, the response function is no longer well defined because
the switching function then has a sharp cut-off at a singularity of
the Wightman distribution. In special cases in which the trajectory is
stationary, the vacuum state is invariant under the Killing vector
generating this stationary motion and the detector has been switched
on in the infinite past
\cite{unruh,DeWitt,gibb-haw:dS,hawking}, the issue can be bypassed by
formally integrating over the whole trajectory and factoring out the
infinite total proper time, because by stationarity the transition
rate can then be argued to be time-independent. But in a general
setting this is not possible, and seemingly inconspicuous
regularisations of the Wightman distribution can lead to unphysical
results, even for uniformly accelerated motion in Minkowski space
\cite{schlicht,profile}.

A way to address this problem is to regard the sharp detector
switch-off as a limit of a family of smooth switch-offs and
investigate how the results depend on the way in which the limit is
taken. In \cite{switching} this issue was investigated for a massless
scalar field in four-dimensional Minkowski space, with the quantum
field in the Minkowski vacuum. The response function with a smooth
switching function was written in a form in which the integrand is no
longer a distribution but a genuine function, and it was shown that a
well-defined notion of a transition rate emerges when the switching
time scale is small compared with the total duration of the
coupling. It was also shown that in the appropriate limits this
transition rate coincides with that obtained by regularising the
sharply switched-off detector by a nonzero spatial size
\cite{schlicht,profile}. The key point is that when the Wightman
distribution under the integral is represented by an
$i\epsilon$-regularised function, the regulator limit
$\epsilon\rightarrow0$ and the limit to sharp switching do not in
general commute and the first must be taken before the second.

The aim of this paper is to extend these results to a more
general setting. For this, we will start in section 
\ref{sec:detectors}
with a review of
the Unruh-DeWitt detector, with special attention to 
the procedure introduced in \cite{switching}
that allows limits of switching functions to be considered.  
In section 
\ref{sec:eps-limit}
the results of
\cite{switching} are generalised to a situation in which Minkowski
space is replaced by an arbitrary four-dimensional globally hyperbolic
spacetime, the Minkowski vacuum state by an arbitrary Hadamard state
and the massless scalar field by a scalar field with arbitrary mass
and curvature coupling. We shall in particular obtain a simple and
manifestly well-defined expression for the difference in the response
of detectors that have the same switching function but move
in different quantum states of the field, on different trajectories 
or even in different spacetimes. 
The limit of sharp switching is
discussed in section~\ref{sec:sharp}. In sections 
\ref{sec:rindler} and 
\ref{sec:newtonian}
we use these results to obtain the detector transition rate in two
examples of interest: an inertial detector in the Rindler vacuum in
Minkowski space, and a detector at rest in a 
static Newtonian gravitational
field. The results are summarised and discussed in
section~\ref{sec:conclusions}. 
Certain technical properties of the detector response in the 
Rindler vacuum are established in the Appendix. 

Throughout this paper we will assume a Lorentzian metric of signature
$({-}{+}{+}{+})$, using the $({+}{+}{+})$ sign 
convention of Misner, Thorne and Wheeler~\cite{mtw}. 
We use units in which 
$c=\hbar=1$, while keeping $G=l_p^2\neq1$. 
Spacetime points are denoted by sans-serif letters.
The symbol $O(x)$ denotes a quantity for which $O(x)/x$ is bounded as
$x\to0$. $O(1)$~denotes a quantity that is bounded in the limit under
consideration.

\section{Particle detectors and their regularisation}
\label{sec:detectors}

We consider a detector consisting of an idealised atom with two energy
levels, $\vert0\rangle_d$ and~$\vert1\rangle_d$, with associated
energy eigenvalues $0$ and $\omega$. The detector is following a
timelike $C^\infty$ trajectory~$\mathsf{x}(\tau)$, parametrised by its
proper time~$\tau$, in a four-dimensional Lorentzian globally
hyperbolic $C^\infty$ manifold~$M$. The coupling of the
detector to a real scalar field $\phi$ of mass $m$ and curvature
coupling $\xi$ is given by the interaction
Hamiltonian
\begin{equation}
\label{Hint}
H_{\mathrm{int}}(\tau)= c \chi(\tau) \mu(\tau)\phi 
\bigl(\mathsf{x}(\tau) \bigr) \,, 
\end{equation}
where $c$ is a coupling constant, $\mu(\tau)$ is the detector's
monopole moment operator and $\chi(\tau)$ is a smooth non-negative
function of compact support. $\chi$~is called the \emph{switching
function\/}: the interaction takes place only when $\chi$ is
nonvanishing, and because $\chi$ has compact support the interaction
has a finite duration. If $\mathsf{x}(\tau)$ is not defined for all
$\tau\in\BbbR$, we assume the support of $\chi(\tau)$ to be in the
open interval in which $\mathsf{x}(\tau)$ is defined.

We take the initial state
of the joint system before the interaction to be $\vert\Psi\rangle
\otimes\vert0\rangle_d$, where the field state $\vert\Psi\rangle$ is
an arbitrary Hadamard state~\cite{kay-wald,radzikowski}. 
We are interested in the probability for
the detector to be observed at state $\vert1\rangle_d$ after the
interaction has been switched off. 
Treating the coupling constant $c$ as a small parameter, 
working to first order in perturbation theory in~$c$, and summing over
the unobserved final state of the field, this probability reads
\cite{Junker,byd,wald-smallbook} 
\begin{equation}
\label{total-probability}
P(\omega)
=
c^2
\, 
{\bigl\vert
{}_d\langle0\vert\mu(0)\vert1\rangle_d\bigr\vert}^2
F(\omega)
\ , 
\end{equation}
where the response function 
$F(\omega)$ is given by 
\begin{equation}
\label{defresponse}
F(\omega)= \int_{-\infty}^{\infty}
\mathrm{d}\tau' \int_{-\infty}^{\infty}\mathrm{d}\tau'' 
\, 
\mathrm{e}^{-i\omega(\tau'-\tau'')}
\, 
\chi(\tau')\chi(\tau'')
\, 
W(\tau',\tau'')
\end{equation} 
and the distributional correlation function $W(\tau',\tau'')$ 
is the pull-back of the Wightman distribution 
$W (\mathsf{x},\mathsf{x}') 
:= \langle \Psi\vert \phi(\mathsf{x})\phi(\mathsf{x}')\vert
\Psi\rangle$
to the detector worldline,\footnote{We denote both the 
spacetime Wightman distribution $W(\mathsf{x},\mathsf{x}')$ 
and its 
pull-back $W(\tau,\tau')$ to the detector worldline by 
the same letter, 
writing out the arguments explicitly in 
places where ambiguity could arise.} 
\begin{equation}
\label{defW} 
W(\tau',\tau'') := 
W \bigl(\mathsf{x}(\tau'),\mathsf{x}(\tau'') \bigr) 
\ . 
\end{equation}
The response function thus encodes the properties that depend on the
state $\vert\Psi\rangle$ and the detector trajectory, while the
prefactor in (\ref{total-probability}) is a constant that only depends
on the detector's internal properties. We shall from now on 
suppress the prefactor and refer to the response function simply 
as the probability. 

To summarise, 
(\ref{defresponse}) gives an unambiguous answer
to the question ``What is the probability of the detector being
observed in the state $\vert1\rangle_d$ after the interaction has
ceased?''

The meaning of the distributional correlation function under the
integral in (\ref{defresponse}) is somewhat subtle. Recall that
the Wightman distribution $W(\mathsf{x},\mathsf{x}')$ 
in a Hadamard state can be
represented by a family of functions \cite{kay-wald,radzikowski}
\begin{equation}
\label{defHadamard}
 W_\epsilon(\mathsf{x},\mathsf{x}')=\frac{1}{{(2\pi)}^2}
 \left[
 \frac{\Delta^{1/2}(\mathsf{x},\mathsf{x}')}
{\sigma_\epsilon(\mathsf{x},\mathsf{x}')}
 +v(\mathsf{x},\mathsf{x}')
\ln \bigl(\sigma_\epsilon(\mathsf{x},\mathsf{x}')\bigr)
 +H(\mathsf{x},\mathsf{x}')
 \right] \,,
\end{equation}
where $\epsilon$ is a positive parameter,
$\sigma(\mathsf{x},\mathsf{x}')$ is the squared geodesic distance
between $\mathsf{x}$ and~$\mathsf{x}'$, 
$\sigma_\epsilon
(\mathsf{x},\mathsf{x}') :=
\sigma (\mathsf{x},\mathsf{x}')
+2i\epsilon[T(\mathsf{x})-T(\mathsf{x}')]+\epsilon^2$ and $T$ is any
globally-defined future-increasing $C^\infty$ function. The logarithm
denotes the branch that is real-valued on the positive real axis and
has the cut on the negative real
axis. $\Delta(\mathsf{x},\mathsf{x}')$ is the Van Vleck determinant,
which is smooth for sufficiently near-by $\mathsf{x}$
and~$\mathsf{x}'$, the function $v(\mathsf{x},\mathsf{x}')$ is a
polynomial in $\sigma(\mathsf{x},\mathsf{x}')$, and the function
$H(\mathsf{x},\mathsf{x}')$ can be chosen $C^n$ for arbitrarily large
$n$ by taking the degree of the polynomial $v(\mathsf{x},\mathsf{x}')$
sufficiently high. The $i\epsilon$-prescription in (\ref{defHadamard})
defines the singular part of $W(\mathsf{x},\mathsf{x}')$: the action
of the Wightman distribution is obtained by integrating $W_\epsilon
(\mathsf{x},\mathsf{x}')$ against test functions and taking the limit
$\epsilon\to0$, and this limit can be shown to be independent of the
choice of the global time function~$T$. Now, the distributional
correlation function $W(\tau,\tau')$ (\ref{defW}) is the pull-back of
$W (\mathsf{x},\mathsf{x}')$ to the detector's worldline, which is a
$C^\infty$ submanifold. It follows that the action of the distribution
$W(\tau,\tau')$ is obtained by pulling back the function
$W_\epsilon(\mathsf{x},\mathsf{x}')$ to the function $W_\epsilon
(\tau,\tau')$, integrating $W_\epsilon (\tau,\tau')$ against test
functions and taking the limit $\epsilon\to0$
\cite{Junker,hormander-vol1,hormander-paper1}. Formula
(\ref{defresponse}) must thus be understood as
\begin{equation}
\label{defresponse-eps}
F(\omega)= \lim_{\epsilon\to0} 
\int_{-\infty}^{\infty}
\mathrm{d}\tau' \int_{-\infty}^{\infty}\mathrm{d}\tau'' 
\, 
\mathrm{e}^{-i\omega(\tau'-\tau'')}
\, 
\chi(\tau')\chi(\tau'')
\, 
W_\epsilon (\tau',\tau'')
\,, 
\end{equation} 
where the integrand is now an ordinary function and the singular part
of $W(\tau,\tau')$ 
has been encoded in the $i\epsilon$ prescription. As
$\overline{W_\epsilon(\mathsf{x},\mathsf{x}')} =
W_\epsilon(\mathsf{x}',\mathsf{x})$, 
where the oveline denotes complex conjugation, 
we have
$\overline{W_\epsilon(\tau',\tau'')} = W_\epsilon(\tau'',\tau')$, and
it follows that (\ref{defresponse-eps}) can be written in the
equivalent form
\cite{schlicht,switching}
\begin{equation}\label{newvariables}
F(\omega)=2\,
\lim_{\epsilon\to0} 
\mathrm{Re}\int_{-\infty}^{\infty}\mathrm{d}u \,
\chi(u)\int_{0}^{\infty}\mathrm{d}s \,\chi(u-s)\, 
\mathrm{e}^{-i\omega s}\, W_\epsilon(u,u-s)\,.
\end{equation}

Although formulas (\ref{defresponse-eps}) and (\ref{newvariables}) are
suitable for computing the detector's response, these formulas do not
display a clear separation between those properties of the response
that depend on on the trajectory and the quantum state and those
properties that only depend on the choice of the switching
function. Neither do these formulas exhibit how the response depends
on the proper time along a given trajectory. Several authors
\cite{schlicht,finitetime1,finitetime2,finitetime3} have therefore
addressed the question: ``If the detector is turned on at proper time
$\tau_0$ and read at proper time~$\tau$, while the interaction is
still on, what is the probability that the transition has taken
place?'' If issues of regularisation could be ignored, this would amount 
to adopting in (\ref{defresponse}) the switching function
\begin{equation}
\label{eq:sharp-switching}
\chi(\tau')=\Theta(\tau'-\tau_0)\Theta(\tau-\tau')
\, , 
\end{equation}
where $\Theta$ is the Heaviside function. 
The transition probability then becomes a function of the reading time 
$\tau$ and can be written as 
\begin{equation}
\label{instant-prob}
F_{\tau}(\omega)=2\,\mathrm{Re}
\int_{\tau_0}^{\tau}\mathrm{d}u\int_0^{u-\tau_0}\mathrm{d}s\,
\mathrm{e}^{-i\omega s}\, W(u,u-s)\,,
\end{equation}
and we can define the 
instantaneous transition rate as its derivative with respect to~$\tau$, 
\begin{equation}
\label{defexcitationrate}
\dot{F}_{\tau}(\omega)=2\,\mathrm{Re}\int_0^{\Delta\tau}\mathrm{d}s\,
\mathrm{e}^{-i\omega s}\, W(\tau,\tau-s)\,,
\end{equation}
where $\Delta\tau := \tau-\tau_0$. $\dot{F}_{\tau}(\omega)$ thus
represents the number of transitions per unit time in an ensemble of
identical detectors. It is this instantaneous transition rate that one
expects to have the $\tau$-independent Planckian spectrum in the Unruh
effect in Minkowski spacetime and in its generalisations to curved
spacetimes, once $\tau_0$ is taken to $-\infty$ to avoid transient
effects.  A~comprehensive recent review of the Unruh effect can be
found in~\cite{crispino-etal}.

We note in passing that the transition rate (\ref{defexcitationrate})
is not directly related to transition rates that could be measured
with a single ensemble of detectors. Given an ensemble of identical
detectors on a given trajectory, $F_\tau(\omega)$ gives the fraction
of detectors that have undergone a transition when observed at
time~$\tau$, but as an observation alters the dynamics of the system,
$F_\tau(\omega)$ no longer has this interpretation after a first
observation has been made. To measure $\dot{F}_\tau(\omega)$, one
therefore needs a set of identical ensembles, such that each ensemble
is used to measure $F_\tau(\omega)$ at just a single value
of~$\tau$. Note in particular that $\dot{F}_\tau(\omega)$ may well be
negative at some values of $\tau$
\cite{Langlois,Langlois-thesis}.  $\dot{F}_\tau(\omega)$ may thus be
difficult to measure operationally, but it is nevertheless of interest
as a nonstationary generalisation of the transition rate that
naturally arises in stationary situations.

Returning to formulas (\ref{instant-prob})
and~(\ref{defexcitationrate}), the difficulty with them as written is
that the `switching function' (\ref{eq:sharp-switching}) is not
smooth. We are no longer guaranteed that replacing $W(\tau,\tau')$ by
$W_\epsilon (\tau,\tau')$ in (\ref{instant-prob}) and
(\ref{defexcitationrate}) and taking the limit $\epsilon\to0$ would
give a result that is independent of the choice of the global time
function in~(\ref{defHadamard}). Case studies have shown that the
result depends on the choice of the time function for Minkowski vacuum
in Minkowski space
\cite{schlicht,profile,Langlois,Langlois-thesis} 
and the Euclidean vacuum in de~Sitter space
\cite{Langlois,Langlois-thesis}, and the methods of Appendix A of
\cite{profile} can be adapted to show that the same holds for
arbitrary Hadamard states in an arbitrary spacetime. Formula
(\ref{defexcitationrate}) does therefore not provide a well-defined
notion of an instantaneous transition rate.

One way to address this problem was introduced in \cite{schlicht} and
further developed in
\cite{profile,Langlois,Langlois-thesis}. The idea is to 
replace the correlation function $W(\tau,\tau')$ 
in (\ref{defexcitationrate}) by a
correlation function in which the field operator has been smeared over
a spacelike hypersurface orthogonal to the trajectory. The weight
function in the smearing is characterised by a positive length
parameter~$\epsilon$, which acts as a regulator and corresponds
physically to the spatial size of the detector in its instantaneous
rest frame. At the end the pointlike detector limit $\epsilon\to0$ is
taken. This scheme does not rely on the choice of a time function to
regularise the Wightman distribution, and in Minkowski space the
introduction of the spatial hypersurfaces is straightforward
\cite{schlicht} and there are partial results regarding independence
of the choice of the weight function~\cite{profile}. An implementation
of the scheme in de~Sitter space was given in
\cite{Langlois,Langlois-thesis,MG}. However, the spacelike surfaces
introduced in Minkowski space in
\cite{schlicht} 
are not easily generalisable to spacetimes without a high degree of
symmetry, and it would seem desirable to attach a meaning to the
instantaneous transition rate within the framework of the conventional
regularisation of the Wightman distribution~(\ref{defHadamard}).

A way that stays fully within the conventional regularisation of
(\ref{defHadamard}), (\ref{defresponse-eps}) and (\ref{newvariables})
was introduced in \cite{switching} in the special case of Minkowski
spacetime, massless scalar field and the Minkowski vacuum
state. Adopting a Lorentz frame with global Minkowski coordinates
$(t,\mathbf{x})$ and choosing $t$ as the global time function, formula
(\ref{newvariables}) for the transition probability becomes
\begin{equation}
\label{switch-Mink}
  F(\omega)=\frac{1}{2\pi^2}\,\lim_{\epsilon\rightarrow 0}
  \,\mathrm{Re}\int_{-\infty}^{\infty}\mathrm{d}u
  \,\chi(u)\int_{0}^{\infty}\mathrm{d}s \,\chi(u-s)\,
  \mathrm{e}^{-i\omega s}\,\,\frac{1}{\left(\Delta \mathsf{x}\right)
  ^2+2i\epsilon\Delta t+\epsilon^2}\,,
\end{equation}
where ${\left(\Delta \mathsf{x}\right)}^2$ is the squared geodesic
distance between $\mathsf{x}(u)$ and $\mathsf{x}(u-s)$ and $\Delta t
:= t(u) - t(u-s)$. The limit $\epsilon\to0$ can be computed
explicitly, with the result \cite{switching}
\begin{align}
\label{prob-Mink}
F(\omega) 
&=
-\frac{\omega}{4\pi}\int_{-\infty}^{\infty}\mathrm{d}u\,[\chi(u)]^2 
\, 
+
\frac{1}{2\pi^2}\int_0^{\infty}\frac{\mathrm{d}s}{s^2}
\int_{-\infty}^{\infty}\mathrm{d}u\,\chi(u)\left[
\chi(u)-\chi(u-s)\right]
\nonumber
\\[1ex] 
& 
\hspace{3ex}
+\frac{1}{2\pi^2}\int_{-\infty}^{\infty}\mathrm{d}u\,\chi(u)
\int_0^{\infty}\mathrm{d}s\,\chi(u-s)\left(
\frac{\cos(\omega s)}{{(\Delta \mathsf{x})}^2}+\frac{1}{s^2}\right)
\,. 
\end{align}
When the switching function $\chi$ equals 1 over an interval of length
$\Delta\tau$, and the switch-on and switch-off each take place within
an interval of length $\delta$ with a profile that scales with
$\delta$ but whose shape is otherwise fixed, the leading behaviour of
the transition rate (defined as the derivative of $F(\omega)$
(\ref{prob-Mink}) with respect to $\Delta\tau$) at $\delta\to0$ is
\begin{equation}
\label{resultado1-sec4}
\dot{F}_{\tau}(\omega)
=
-\frac{\omega}{4\pi}+\frac{1}{2\pi^2}
\int_0^{\Delta\tau}\textrm{d}s
\left( 
\frac{\cos (\omega s)}{{(\Delta \mathsf{x})}^2} 
+ 
\frac{1}{s^2} 
\right) 
\ \ +\frac{1}{2\pi^2 \Delta \tau} 
+ O(\delta) 
\, , 
\end{equation}
where now $\left(\Delta \mathsf{x}\right) ^2$ is the squared geodesic
distance between $\mathsf{x}(\tau)$ and $\mathsf{x}(\tau-s)$. In the
limit $\delta\to0$, the transition rate (\ref{resultado1-sec4}) agrees
with that obtained from spatial smearing in~\cite{profile}, and it
reproduces the expected Planckian spectrum when the trajectory is
uniformly linearly accelerated and $\Delta\tau\rightarrow
\infty$. Further properties of this transition rate are discussed in
\cite{profile,switching}.

In this paper we generalise the Minkowski vacuum results
(\ref{prob-Mink}) and (\ref{resultado1-sec4}) to a general Hadamard
vacuum state in four-dimensional spacetime, 
for a field with arbitrary values of the mass and the curvature coupling. 
We shall show that most of the arguments in
\cite{switching} carry over to this situation, and we
shall find the expressions that generalise (\ref{prob-Mink})
and~(\ref{resultado1-sec4}). These expressions will then be applied to
two examples.

\section{Regulator-free response function in 
a general Hadamard state}
\label{sec:eps-limit}


In this section we obtain a regulator-free expression for the response
function $F(\omega)$ by computing explicitly the limit $\epsilon\to0$
in~(\ref{newvariables}). Following the procedure used
in~\cite{switching}, we split the $s$-integral into the subintervals
$(0,\eta)$ and $(\eta,\infty)$, with $\eta=\sqrt{\epsilon}$, estimate
the integrand in each subinterval and finally combine the results.

We shall make use of the small $s$ expansions
\begin{subequations}
\label{eq:small-expansions}
\begin{align}
\label{eq:sigma-small-expansion}
\sigma & = -s^2-\frac{1}{12}a^2s^4+O(s^5) 
\,, 
\\
\label{eq:Delta-small-expansion}
\Delta &= 1+O(s^2)
\,, 
\\
\label{eq:v-small-expansion}
v &= m^2+\left( \xi-\frac{1}{6}\right) R+O(s^2)
\,, 
\\
\label{eq:DeltaT-small-expansion}
\Delta T &=\dot{T}s-\frac{\ddot{T}s^2}{2}+O(s^3)
\,, 
\end{align}
\end{subequations} 
where the Ricci scalar~$R$, the squared (covariant) acceleration
$a^2$ and $T$ are evaluated at the point $\mathsf{x}(u)$ and the dots 
indicate proper time derivatives.

\subsection{Subinterval $s\in(\eta,\infty)$} 
\label{subsec:large}

Consider in (\ref{newvariables}) the subinterval
$s\in(\eta,\infty)$, and let 
$W_0$ denote the pointwise limit of $W_\epsilon$ as $\epsilon\to0$. 
Replacing $W_\epsilon$ by 
$W_0$ 
creates under the $u$-integral an error that 
equals $\chi(u)$ times the quantity 
\begin{align}
\label{error}
 & 2\,\mathrm{Re} \int_\eta^{\infty} \mathrm{d}s
 \,\chi(u-s)\,\mathrm{e}^{-i\omega s}\bigl[
 W_{\epsilon}(u,u-s)-W_0(u,u-s)\bigr] \nonumber\\ & =
 \frac{1}{2\pi^2}\,\mathrm{Re}\int_\eta^{\infty} \mathrm{d}s
 \,\chi(u-s)\,\mathrm{e}^{-i\omega s} \left\{ \Delta^{1/2}\left(
 \frac{1}{\sigma_\epsilon}-\frac{1}{\sigma}\right)+v\bigl[
 \ln(\sigma_\epsilon)-\ln(\sigma)\bigr] \right\} \,,
\end{align}
where the functions $\Delta$, $v$, $\sigma$ and $\sigma_\epsilon$ are
each evaluated at the pair
$(\mathsf{x},\mathsf{x}')=\bigl(\mathsf{x}(u),\mathsf{x}(u-s)\bigr)$
and $\ln(\sigma) := \lim_{\epsilon\to0_+} \ln(\sigma_\epsilon)$. 
We shall show that this error term does not contribute to 
(\ref{newvariables}) after the limit $\epsilon\to0$ is taken. 

Consider first in (\ref{error}) the contribution from
$\sigma_\epsilon^{-1}$ and~$\sigma^{-1}$. We split this contribution
into its odd and even parts in~$\omega$. The part that is odd in
$\omega$ can be written as
\begin{align}
\label{upper-odd}
-\frac{1}{2\pi^2}\int_\eta^{\infty} \mathrm{d}s \,
\chi(u-s)\,\sin(\omega s)\Delta^{1/2}
\frac{2\epsilon\Delta T}{\displaystyle \sigma^2
\left[ \left( 1+\frac{\epsilon^2}{\sigma}\right)^2
+\frac{4\epsilon^2{(\Delta T)}^2}{\sigma^2} \right] }
\,, 
\end{align}
where $\Delta T :=
T\bigl(\mathsf{x}(u)\bigr)-T\bigl(\mathsf{x}(u-s)\bigr)$.  
As the switching function makes the upper limit of the 
$s$-integral finite, it follows from (\ref{eq:sigma-small-expansion}) 
and (\ref{eq:DeltaT-small-expansion}) that 
the quantities ${(\Delta T)}^2/\sigma$ and $\epsilon/\sigma$ are bounded
by constants that are independent of~$\eta$.  
We can therefore write (\ref{upper-odd}) as
\begin{equation}
\label{upper-odd2}
-\frac{1}{\pi^2}
\int_\eta^{\infty} \mathrm{d}s \,
\chi(u-s)\,\Delta^{1/2}\,
\frac{\epsilon \sin(\omega s)\Delta T}{\sigma^2} 
\bigl[ 1 + O(\epsilon) \bigr] \,,  
\end{equation}
where the $O(\epsilon)$ estimate holds uniformly in~$s$. 
As the functions $\chi$ and $\Delta$ are $O(1)$ at small~$s$, the
integrand in (\ref{upper-odd2}) is bounded by a constant
times~$\epsilon \, s^{-2}$, 
and (\ref{upper-odd2}) is thus of order $O(\epsilon
\,\eta^{-1})=O(\eta)$. 
Similarly, the part that is even in $\omega$ 
can be written as 
\begin{align}
\label{upper-even}
-\frac{1}{2\pi^2}\int_\eta^{\infty} \mathrm{d}s \,
\chi(u-s)\,\cos(\omega
s)\Delta^{1/2}\,\frac{\epsilon^2}{\sigma^2}\,
\frac{\displaystyle 1+\frac{\epsilon^2}{\sigma}-\frac{4{(\Delta
T)}^2}{\sigma}}{\displaystyle \left( 1+\frac{\epsilon^2}{\sigma}\right)^2+
\frac{4\epsilon^2{(\Delta T)}^2}{\sigma^2}}\, , 
\end{align}
and similar estimates show that the integrand in (\ref{upper-even}) is
bounded by a constant times~$\epsilon^2 s^{-4}$. 
The expression (\ref{upper-even}) is
hence of order $O(\epsilon^2 \,\eta^{-3})=O(\eta)$.

Consider then in (\ref{error}) the contribution from the logarithmic
terms. Keeping track of the branches of the logarithms, we can write
this contribution as
\begin{equation}
\label{eq:upper-logs} 
\frac{1}{2\pi^2}\,\mathrm{Re} 
\int_\eta^{\infty} \mathrm{d}s \,
\chi(u-s)\,\mathrm{e}^{-i\omega s}\,v\,
\ln\left( 
1+\frac{2i\epsilon\Delta T}{\sigma}+\frac{\epsilon^2}{\sigma}
\right) \,.
\end{equation}
It follows from (\ref{eq:sigma-small-expansion}) 
and (\ref{eq:DeltaT-small-expansion}) that 
$\epsilon\Delta T/\sigma$ is bounded by $\eta$ times a constant and
$\epsilon^2/\sigma$ is bounded by $\epsilon$ times a constant. 
The logarithm is hence of order $O(\eta)$ uniformly in~$s$,
and (\ref{eq:upper-logs}) is of order~$O(\eta)$.

As $\chi$ has compact support, 
all the estimates above hold uniformly in $u$ 
under the $u$-integral in~(\ref{newvariables}).
In the subinterval $s\in(\eta,\infty)$ in~(\ref{newvariables}), 
$W_\epsilon$ can
therefore be replaced by $W_0$ without error.

\subsection{Subinterval $s\in(0,\eta)$} 
\label{subsec:small}

We now turn to the subinterval $s \in (0,\eta)$
in~(\ref{newvariables}). The singularity of $W$ at $s=0$ implies that
we cannot directly replace $W_\epsilon$ by~$W_0$, and we will need to
examine the small $s$ behaviour of $W_\epsilon$ more closely. 


We observe first that the term $H(\mathsf{x},\mathsf{x}')$ in
(\ref{defHadamard}) clearly gives a vanishing contribution
to~(\ref{newvariables}).

Consider then the logarithmic term in~(\ref{defHadamard}). Suppressing
for the moment the factor~$\chi(u)$, the integral over $u$ and the
limit $\epsilon\to0$, the contribution to (\ref{newvariables}) reads
\begin{align}
\label{eq:small-log}
\frac{1}{2\pi^2}\,\mathrm{Re}\int_0^\eta\mathrm{d}s\,
\chi(u-s)\mathrm{e}^{-i\omega s}
\left\{ 
\left[ m^2+\left( \xi-\frac{1}{6}\right) R\right] 
+O(s^2)
\right\} 
\ln\bigl(\sigma_\epsilon \bigr) \,. 
\end{align}
The imaginary part of the logarithm is bounded and its contribution in 
(\ref{eq:small-log}) is therefore of order~$O(\eta)$. 
To estimate the real part of the logarithm, 
we write $s = \epsilon x$, with $0<x<1/\eta$, and use the expansions 
(\ref{eq:small-expansions}) to obtain 
\begin{align}
\label{eq:modsigmaeps-estimate}
{\bigl| \sigma_\epsilon \bigr|}^2 
&= 
\epsilon^4 
\left[ {(1-x^2)}^2 +4 x^2 {\dot{T}}^2 \right] 
\bigl[ 1 + O(\eta) \bigr] 
\,, 
\end{align} 
where the $O(\eta)$ term holds uniformly in~$x$. Hence
\begin{align}
2 \, \mathrm{Re} 
\ln\bigl(\sigma_\epsilon\bigr) 
&= 
4 \ln\epsilon 
+ 
\ln 
\left[ {(1-x^2)}^2 +4 x^2 {\dot{T}}^2 \right] 
\ + O(\eta)
\,, 
\end{align} 
where again the $O(\eta)$ term holds uniformly in~$x$. The
contribution in (\ref{eq:small-log}) is therefore of order $O(\eta
\ln\eta)$. As this estimate holds uniformly in $u$, by virtue of the
compact support of~$\chi$, the logarithmic term does thus not
contribute in~(\ref{newvariables}).

Finally, consider the $\sigma_\epsilon^{-1}$ term
in~(\ref{defHadamard}). From (\ref{eq:modsigmaeps-estimate}) we see
that $s^2 \, \sigma_\epsilon^{-1}$ is bounded, and hence
$\int_0^\eta\mathrm{d}s\, s^2 \, \sigma_\epsilon^{-1}\,=O(\eta)$. It
follows from (\ref{eq:Delta-small-expansion}) that we may replace
$\Delta^{1/2}$ by~$1$, and we may similarly replace the factor 
$\chi(u-s) \mathrm{e}^{-i\omega s}$ by 
$(1- i\omega s)\chi(u) - s \dot{\chi}(u)$. 
What remains is to analyse 
the small $\eta$
behaviour of the expression
\begin{equation}
\label{eq:Ismall-def}
I_< := \frac{1}{2\pi^2} \, 
\mathrm{Re}\int_0^{\eta}\mathrm{d}s\,
\frac{(1- i\omega s)\chi - s \dot{\chi}}{\sigma_\epsilon}
\,, 
\end{equation}
where $\chi$ and $\dot{\chi}$ are evaluated at~$u$. 
In the special case in which the spacetime is Minkowski space and the
global time function $T$ is the Minkowski time coordinate in a given
Lorentz frame, this analysis was carried out in~\cite{switching}, and
the techniques used therein generalise to (\ref{eq:Ismall-def}) in a
straightforward way. Splitting $I_<$ into its even and odd parts in
$\omega$ as $I_< = I_<^{\mathrm{even}} + I_<^{\mathrm{odd}}$, and
writing $s = \epsilon x$ with $0<x< 1/\eta$, 
we find\footnote{Our formula (\ref{expodd}) 
corrects a typographical error in formula (3.4b) of~\cite{switching}.} 
\begin{subequations}
\label{expansion}
\begin{align}
\label{expeven}
I_<^{\mathrm{even}} &=
\frac{1}{2\pi^2}\int_0^{1/\eta}
\frac{(1-x^2) \,\mathrm{d}x}
{{(1-x^2)}^2 +4 x^2 {\dot{T}}^2}
\left[
\frac{\chi}{\eta^2}-\dot{\chi} x 
+ \frac{4\chi\dot{T}\ddot{T} x^3}{{(1-x^2)}^2 +4 x^2 {\dot{T}}^2}
\right] 
\ \ \ +O(\eta) 
\,, 
\\
\label{expodd}
I_<^{\mathrm{odd}} &=
-\frac{\omega \chi \dot{T}}{\pi^2}
\int_0^{1/\eta}
\frac{x^2 \, \mathrm{d}x}{{(1-x^2)}^2 +4 x^2 {\dot{T}}^2}
\ \ \ +O(\eta)
\,.
\end{align}
\end{subequations}
In (\ref{expeven}) the integral of the first term is elementary, and
multiplying the second and third term by $\chi$ yields a total
$u$-derivative that can be taken outside the integral. The result is
\begin{align}
\label{expeven-more}
\chi I_<^{\mathrm{even}} &=
\frac{\chi^2}{2\pi^2 \eta}
-
\frac{1}{4\pi^2}
\frac{\mathrm{d}}{\mathrm{d}u}
\int_0^{1/\eta}
\frac{\chi^2 \, x(1-x^2) \, \mathrm{d}x}{{(1-x^2)}^2 +4 x^2 {\dot{T}}^2}
\ \ \ +O(\eta)
\,.
\end{align}
The integral in (\ref{expodd}) is elementary, 
and multiplying the result by $\chi$ we obtain 
\begin{align}
\label{expodd-more}
\chi I_<^{\mathrm{odd}} &=
-\frac{\omega \chi^2 }{4\pi}
+O(\eta)
\,.
\end{align}
All these estimates hold uniformly in~$u$, 
owing to the compact support of~$\chi$. The only terms that contribute in 
the subinterval 
$s \in (0,\eta)$
in (\ref{newvariables}) are therefore the explicitly-displayed terms in 
(\ref{expeven-more}) and~(\ref{expodd-more}).

\subsection{Joining the subintervals} 

Substituting the results of 
subsections 
\ref{subsec:large}
and 
\ref{subsec:small} in~(\ref{newvariables}), we find 
\begin{align}
F(\omega) &=
-\frac{\omega}{4\pi}\int_{-\infty}^{\infty}\mathrm{d}u\,{[\chi(u)]}^2 
\nonumber 
\\
&\hspace{3ex}
+ 
2 \lim_{\eta\rightarrow 0} 
\int_{-\infty}^{\infty}\mathrm{d}u 
\, 
\chi(u)
\left[ 
\frac{\chi(u)}{4 \pi^2 \eta}
+
\mathrm{Re}
\int_\eta^{\infty}\mathrm{d}s\,\chi(u-s)\,
\mathrm{e}^{-i\omega s}\,W_0(u,u-s)
\right]
\,.  
\label{eq:Flim3}
\end{align}
As $\chi$ has compact support, the total derivative term in 
(\ref{expeven-more}) integrates to zero and has dropped out. 
What remains is to take the limit in~(\ref{eq:Flim3}). 

Following \cite{switching}, we take the term proportional to 
$1/\eta$ under the $s$-integral, add and subtract under the $s$-integral 
the term $\chi(u-s)/(4\pi^2 s^2)$ and group the terms in the form 
\begin{align}
F(\omega) 
&=
-\frac{\omega}{4\pi}\int_{-\infty}^{\infty}\mathrm{d}u\,{[\chi(u)]}^2 
\nonumber 
\\
&\hspace{3ex}
+ 
2 
\lim_{\eta\rightarrow 0} 
\int_{-\infty}^{\infty}\mathrm{d}u 
\, 
\chi(u)
\int_\eta^{\infty}\mathrm{d}s
\, \mathrm{Re} 
\left[
\chi(u-s)\,
\mathrm{e}^{-i\omega s}\,W_0(u,u-s)
+\frac{\chi(u)}{4\pi^2s^2}
\right]
\nonumber
\\
&=
-\frac{\omega}{4\pi}\int_{-\infty}^{\infty}\mathrm{d}u\,{[\chi(u)]}^2 
\nonumber 
\\
&\hspace{3ex}
+ \lim_{\eta\rightarrow 0}\Bigg\{
2
\int_{-\infty}^{\infty}\mathrm{d}u\,\chi(u)
\int_\eta^{\infty}\mathrm{d}s\,\chi(u-s) 
\,\mathrm{Re} 
\left[ 
\mathrm{e}^{-i\omega s}\, W_0(u,u-s)+\frac{1}{4\pi^2s^2}
\right] 
\nonumber
\\
&\hspace{13ex}
+\frac{1}{2\pi^2}\int_\eta^{\infty}
\frac{\mathrm{d}s}{s^2}\int_{-\infty}^{\infty}\mathrm{d}u\,\chi(u)
\bigl[ \chi(u)-\chi(u-s)\bigr] \Bigg\}\,, 
\label{eq:Flim4}
\end{align}
where in the last term the interchange of the $u$-integral and the
$s$-integral is justified by absolute convergence of the double
integral. The limit $\eta\to0$ can now be taken by simply setting
$\eta=0$. In the last term the reason is that the $u$-integral, when
regarded as a function of~$s$, has a Taylor expansion that starts with
$O(s^2)$. In the term involving~$W_0$, the reason is that the real
part of $\mathrm{e}^{-i\omega s}\, W_0(u,u-s)$ has the small $s$
behaviour of $-1/(4\pi^2 s^2)$ plus an integrable function of~$s$, by
virtue of (\ref{defHadamard})
and~(\ref{eq:sigma-small-expansion}). The final result for the
response function is thus
\begin{align}
F(\omega) 
&=
-\frac{\omega}{4\pi}\int_{-\infty}^{\infty}\mathrm{d}u\,{[\chi(u)]}^2 
\ + \ 
\frac{1}{2\pi^2}\int_0^{\infty}
\frac{\mathrm{d}s}{s^2}\int_{-\infty}^{\infty}\mathrm{d}u\,\chi(u)
\bigl[ \chi(u)-\chi(u-s)\bigr] 
\nonumber 
\\
\noalign{\medskip}
&\hspace{3ex}
+ 
2
\int_{-\infty}^{\infty}\mathrm{d}u\,\chi(u)
\int_0^{\infty}\mathrm{d}s\,\chi(u-s) 
\,\mathrm{Re} \left[ 
\mathrm{e}^{-i\omega s}\, W_0(u,u-s)+\frac{1}{4\pi^2s^2}
\right] 
\,. 
\label{probability}
\end{align}
In the special case of the Minkowski vacuum in Minkowski space, 
(\ref{probability}) duly reduces to the 
expression (\ref{prob-Mink}) found in~\cite{switching}. 

The first two terms in (\ref{probability}) depend only on the
switching function $\chi$ but neither on the quantum state, the
spacetime or the trajectory. If we compare two detectors in different
quantum states of the field, on different trajectories or even in
different spacetimes, but having the same switching function, the
difference of the responses is given by
\begin{equation}
\label{diff-response}
\Delta F(\omega)=
2\,\mathrm{Re}\int_{-\infty}^{\infty}\mathrm{d}u\,
\chi(u) 
\int_0^{\infty}\mathrm{d}s\,\chi(u-s)\,
\mathrm{e}^{-i\omega s}\left[ W_0^A(u,u-s)-W_0^B(u,u-s)\right] \,,
\end{equation}
where $W_0^A$ and $W_0^B$ are the pull-backs of the unregularised
Wightman distributions in the two situations. The representation
(\ref{defHadamard}) of the Wightman distribution in a Hadamard state
guarantees that the divergences in (\ref{diff-response}) cancel and
the integral is well defined. This is
particularly convenient for numerical calculations.

\section{Sharp switching limit} 
\label{sec:sharp} 

In this section we discuss the response function (\ref{probability})
in the limit of sharp switch-on and switch-off. As in the case of
Minkowski vacuum~\cite{switching}, we shall isolate the divergence due
to the sharp switching from a finite remainder and show that a
well-defined notion of instantaneous transition rate can be defined in
an appropriate limit.

To control the switch-on and switch-off, we 
assume the switching function to have the form\footnote{Our formula 
(\ref{chihs}) corrects a typographical error 
in the argument of $h_1$ in equation 
(4.1) of~\cite{switching}.}
\begin{equation}
\label{chihs}
\chi(u)=h_1\left( \frac{u-\tau_0+\delta}{\delta}\right)\times h_2
\left( \frac{-u+\tau+\delta}{\delta}\right) \,, 
\end{equation}
where $\tau$, $\tau_0$ and $\delta$ are parameters satisfying 
$\tau_0 < \tau$ and $0<\delta$, and $h_i$, $i=1,2$, 
are non-negative 
$C^\infty$ functions satisfying 
$h_i(x)=0$ for $x\le0$ and 
$h_i(x)=1$ for $1\le x$. 
This means that the detector is turned on smoothly during the interval 
$(\tau_0-\delta,\tau_0)$, with a profile determined by the function~$h_1$, 
it then remains turned on at constant coupling strength 
for the time $\Delta\tau := \tau-\tau_0$, and it is
finally turned off smoothly during the
interval $(\tau,\tau+\delta)$, 
with a profile determined by the function~$h_2$. 
The functions $h_i$ are regarded as fixed. 
We initially regard $\tau_0$ and $\tau$ as fixed but will eventually 
allow $\tau$ to vary. 

The first term in (\ref{probability}) is equal to 
$-(\omega/4\pi) (\Delta\tau + \delta C_1)$, 
where $C_1$ is a positive constant. 
The second term in (\ref{probability}) was analysed 
in~\cite{switching}, 
with the result that it only depends on 
$\Delta\tau$ and $\delta$ through the combination 
$\delta/\Delta\tau$ and has at small 
$\delta/\Delta\tau$ the asymptotic form 
\begin{align}
\label{eq:logterm-as}
\frac{1}{2\pi^2} \ln\left(\frac{\Delta\tau}{\delta}\right) 
+ C_2 
+ O \left(\frac{\delta}{\Delta\tau}\right) 
\,,
\end{align}
where $C_2$ is a constant and the full 
expansion of the $O$-term proceeds 
in positive powers of $\delta/\Delta\tau$.
The last term in (\ref{probability}) can be analysed 
by breaking the integrations into the various 
subintervals as in~\cite{switching}, with the result 
\begin{equation}
\label{eq:plateu-contribution}
2\int_{\tau_0}^{\tau}\mathrm{d}u\,\int_0^{u-\tau_0}
\mathrm{d}s
\,\mathrm{Re}
\left[ 
\mathrm{e}^{-i\omega s}\, W_0(u,u-s)+\frac{1}{4\pi^2s^2}
\right] 
\ \ \ +O(\delta)
\,.
\end{equation}
The qualitatively new feature compared with \cite{switching} is the
logarithmic singularity in $W_0$, but the contribution from this
singularity can be verified to be of order $O(\delta^2\ln\delta)$ and
hence subleading in~(\ref{eq:plateu-contribution}). Collecting, we
find
\begin{align}
F(\omega) & =
-\frac{\omega}{4\pi}\Delta\tau
+2 \int_{\tau_0}^{\tau}\mathrm{d}u\,\int_0^{u-\tau_0}\mathrm{d}s\, 
\,\mathrm{Re} 
\left[ 
\mathrm{e}^{-i\omega s}W_0(u,u-s)+\frac{1}{4\pi^2s^2}
\right]
\nonumber\\
& + \frac{1}{2\pi^2}\ln\left( \frac{\Delta\tau}{\delta}\right) 
+ C_2 +O(\delta) 
\,. 
\label{eq:F-final} 
\end{align}
By the smoothness of the spacetime and the trajectory, we may
differentiate the trajectory-dependent contribution
(\ref{eq:plateu-contribution}) with respect to $\tau$ termwise, and
the same holds for the trajectory-independent contributions by their
explicit structure. We may therefore take the $\tau$-derivative in
(\ref{eq:F-final}) termwise, with the result
\begin{equation}
\label{resultado1}
\dot{F}_{\tau}(\omega)
=
-\frac{\omega}{4\pi}
+2\int_0^{\Delta\tau}\mathrm{d}s
\, \mathrm{Re}
\left[ \mathrm{e}^{-i\omega s}W_0(\tau,\tau-s)+\frac{1}{4\pi^2s^2}\right]
\ \ +\frac{1}{2\pi^2 \Delta \tau} + O(\delta)  
\, . 
\end{equation}

Equations (\ref{eq:F-final}) and (\ref{resultado1}) are our main
result. The transition probability (\ref{eq:F-final}) diverges as
$\delta\to0$, but the divergence has been isolated into an explicit
logarithmic term that is independent of the trajectory or the quantum
state of the field. The $\tau$-derivative of the transition
probability is given by (\ref{resultado1}) and remains finite as
$\delta\to0$. Equation (\ref{resultado1}) provides a definition of
what is meant by the detector's transition rate, without the need to
introduce spatial profiles or other regulators. In the special case of
a massless field in Minkowski spacetime, in the Minkowski vacuum, the
limit $\delta\to0$ in (\ref{resultado1}) recovers the transition rate
obtained in \cite{profile} via a spatial profile regularisation.

We end the section with two comments. First, if we interpret the naive
expression (\ref{defexcitationrate}) for the transition rate as
\begin{equation}
\label{eq:defexcitationrate-epslimit}
\dot{F}_{\tau}(\omega)
=
\lim_{\epsilon\rightarrow 0}
2\,\mathrm{Re}
\int_0^{\Delta\tau}\mathrm{d}s\,
\mathrm{e}^{-i\omega s}\, W_{\epsilon}(\tau,\tau-s) 
\end{equation}
and apply the methods of section~\ref{sec:eps-limit}, we find that
(\ref{eq:defexcitationrate-epslimit}) is equal to the $\delta\to0$
limit of our transition rate (\ref{resultado1}) plus an additional
term proportional to $\ddot T$. The additional term vanishes if the
pull-back of $T$ to the trajectory is an affine function
of~$\tau$. The transition rate of a sharply-switched detector can thus
be calculated equivalently from the $\delta\to0$ limit
in~(\ref{resultado1}), where the singularity in the Wightman
distribution is cancelled by an explicit counterterm, or from the
$\epsilon\to0$ limit in~(\ref{eq:defexcitationrate-epslimit}),
provided the time time function used to regularise
(\ref{eq:defexcitationrate-epslimit}) is an affine function of $\tau$
on the trajectory.

Second, if we compare two detectors in different quantum states of the
field, on different trajectories or even in different spacetimes, but
having been in operation for the same length of proper time, we
recover for the difference of the transition rates in the $\delta\to0$
limit the formula
\begin{equation}
\label{diff-rate}
\Delta \dot{F}_{\tau}(\omega) 
= 
2\, \mathrm{Re}
\int_0^{\Delta\tau}\mathrm{d}s\,\mathrm{e}^{-i\omega s}
\left[
W_0^A(\tau,\tau-s)-W_0^B(\tau,\tau-s)
\right]\,. 
\end{equation}
The difference in the transition rates in two given situations can
thus be written as a Fourier transform of a function that requires no
regularisation. Formula (\ref{diff-rate}) is useful for both
analytical and numerical calculations, especially in cases where the
transition rate in one of the two situations is already known. In the
next two sections we shall apply this formula to two such examples.


\section{Inertial detector in the Rindler vacuum}
\label{sec:rindler}

In this section we consider a detector moving inertially through the
Rindler wedge in Minkowski space, coupled to a massless scalar field
in its Rindler vacuum state. This is the state in which the uniformly
accelerated detectors associated with the Rindler wedge do not get
excited. A~naive application of the equivalence principle could be
argued to imply that as an accelerated detector moving through the
`unaccelerated' (Minkowski) vacuum state gets excited thermally, an
unaccelerated detector moving through the `accelerated' (Rindler)
vacuum should also get excited thermally. Working in the limit
$\delta\to0$, we shall show that this does not hold: the detector does
have a nontrivial transition rate, but the rate is neither thermal nor
constant in the detector's proper time, and it diverges as the
detector approaches the Rindler horizon.


Let $(t,x,y,z)$ be a set of standard Minkowski 
coordinates in Minkowski space. We take the detector to move in the 
`right-hand-side' Rindler wedge, $x>\vert t\vert$, denoted by~$R$. 
The Rindler vacuum Wightman distribution 
$W^R(\mathsf{x},\mathsf{x}')$ in $R$ 
reads \cite{candelas}
\begin{equation}
\label{WR}
W^R(\mathsf{x},\mathsf{x}')
=
W^M(\mathsf{x},\mathsf{x}')-\int_{-\infty}^{\infty} 
\frac{\mathrm{d}v}{\pi^2+v^2} \, 
W^M \bigl(\mathsf{x},\mathsf{x}''(v)\bigr)
\,, 
\end{equation}
where $W^M(\mathsf{x},\mathsf{y}) = 
{[4\pi^2{(\mathsf{x}- \mathsf{y})}^2]}^{-1}$ 
is the Minkowski vacuum Wightman distribution, 
the points $\mathsf{x} = (t,x,y,z)$ and $\mathsf{x}' = (t',x',y',z')$ 
are in $R$ 
and 
$\mathsf{x}''(v) := 
\bigl(
- t'\cosh v - x' \sinh v, 
- x'\cosh v -  t' \sinh v, 
y',z'\bigr)$. 
The difference of 
$W^R(\mathsf{x},\mathsf{x}')$ and 
$W^M(\mathsf{x},\mathsf{x}')$ consists thus of the integral of 
$W^M$ over an orbit of the associated Killing vector in the 
\emph{opposite\/} Rindler wedge. 
Note that since the two Rindler wedges are spacelike separated, the
difference term is a nonsingular function, and the distributional
character of $W^R(\mathsf{x},\mathsf{x}')$ comes entirely from the
first term on the right-hand side in~(\ref{WR}). As we shall be using
formula~(\ref{diff-rate}), we are suppressing the distributional
issues in~(\ref{WR}).

We take the trajectory of the detector to be
\begin{align}
\label{trajectoryR}
\mathsf{x}(\tau)
&=(\tau,X,0,0)
\,,
\end{align}
where $X$ is a positive constant and $\tau$ is the proper time. 
The trajectory stays in $R$ during the proper time interval 
$\vert\tau\vert<X$. 
We must therefore consider the detector response 
in a finite proper time interval. 

We shall compute the transition rate from formula~(\ref{diff-rate}),
using the Minkowski vacuum as reference state. Unlike the infinite
$\Delta\tau$ case, in which the excitation rate is zero, a detector
moving inertially through Minkowski vacuum over a finite proper time
$\Delta\tau$ does have transient excitations due to the
switching. This transition rate has been found in several papers (see
e.g.~\cite{finitetime1}) and equals
\begin{equation}
\label{inert-Mink}
\dot{F}_{\Delta\tau}^{M}(\omega)
=
-\frac{\omega}{4\pi}
+\frac{\cos(\omega \,\Delta\tau)}{2\pi^2\Delta\tau}
+\frac{1}{2\pi^2}\,\omega\,\mathrm{Si}(\omega\,\Delta\tau)
\end{equation}
where $\mathrm{Si}$ is the sine integral function. 
$\dot{F}_{\Delta\tau}^{M}(\omega)$
diverges as $\Delta\tau \rightarrow 0$ 
(which is an artifact of omitting the $O(\delta)$ 
terms in~(\ref{resultado1}))
and approaches 
$[\omega/(2\pi)]\Theta(-\omega)$ for large~$\Delta\tau$.

Using (\ref{diff-rate}) with~(\ref{WR}), 
and substituting the trajectory (\ref{trajectoryR})
with $-X < \tau_0 < \tau < X$, 
the transition rate in the Rindler 
vacuum at time $\tau$ can be written as 
\begin{align}
\dot{F}_{\tau}^{R}(\omega)
&=
\dot{F}_{\Delta\tau}^{M}(\omega)
+ 
\Delta \dot{F}_{\tau}(\omega)
\,,
\end{align}
where 
\begin{align}
\Delta \dot{F}_{\tau}(\omega)
&= 
-\frac{1}{2\pi^2}
\int_0^{\Delta\tau}\mathrm{d}s \,
\cos(\omega s)
\int_{-\infty}^{\infty}
\frac{\mathrm{d}v}{\pi^2+v^2}\nonumber\\
&\ \ \ \times\frac{1}
{-{[\tau+(\tau-s)\cosh v+X\sinh v]}^2
+{[X+X \cosh v+(\tau-s)\sinh v]}^2}
\,. 
\label{eq:log-div-def}
\end{align}
The $v$-integral in (\ref{eq:log-div-def})
can be done by contour integration. 
Closing the contour in the upper half-plane, there are 
two infinite series of contributing poles, at respectively  
$v=\ln(X+\tau)-\ln(X+\tau-s)+(2m+1)i\pi$ 
and 
$v=\ln(X-\tau+s)-\ln(X-\tau)+(2m+1)i\pi$ with $m=0,1,2,\ldots$, 
and a single contributing 
pole at $v=i\pi$. Summing over the residues, we find 
\begin{align}
\label{log-div}
\Delta \dot{F}_{\tau}(\omega)
&=
\frac{1}{2\pi^2}\int_0^{\Delta\tau}\mathrm{d}s \,
\frac{\cos(\omega s)}{s^2}
\left[
1-\frac{s}{2\tau-s}
\left(
\frac{1}{\ln\left(\frac{ X-\tau}{X-\tau+s}\right)}
+\frac{1}{\ln\left(\frac{ X+\tau}{X+\tau-s}\right)}
\right) 
\right] 
\,.
\end{align}
Note that the integrand in (\ref{log-div}) remains finite as $s\to0$,
and the integral is well defined. Regarding $\tau_0$ fixed and $\tau$
as variable, we see that $\Delta \dot{F}_{\tau}(\omega)$ tends to zero
as $\tau\to\tau_0$, but it diverges to $-\infty$ as $\tau\to X$, which
is the limit in which the trajectory approaches the Rindler
horizon. We shall show in the Appendix that the asymptotic form of
$\Delta \dot{F}_{\tau}(\omega)$ at $\tau\to X$ is
\begin{equation}
\label{explicit-logdiv-text}
\Delta \dot{F}_{\tau}(\omega)
=
\frac{1}{2\pi^2X}
\left\{
\frac{1}{4} \ln\left(1 - \frac{\tau}{X}  \right)
+
\frac{1}{2} \ln\left[ -\ln \left(1 - \frac{\tau}{X}  \right) \right] 
+O(1)
\right\}
\,, 
\end{equation}
and the divergence is thus logarithmic in~$\tau$. 

The response of inertial detectors in the Rindler vacuum was
previously studied by Candelas and Sciama~\cite{cande-sciama}, but in
a somewhat different framework. Candelas and Sciama investigate the
whole family of trajectories~(\ref{trajectoryR}), and they compute the
transition rate on each trajectory at the point $\tau = \sqrt{X^2 -
a^{-2}}$, where $a$ is a fixed positive constant. This means that the
inertial trajectories are being compared along a Rindler trajectory of
acceleration~$a$. In the limit $X\to\infty$, it is found that the
response approaches the Minkowski vacuum value
$[\omega/(2\pi)]\Theta(-\omega)$. Candelas and Sciama's interpretation
of this result is that ``for the case of a charge moving inertially in
Minkowski space-time through an \emph{accelerated\/} vacuum the
spectrum of field fluctuations perceived by the charge
\emph{is the same as if the vacuum were unaccelerated\/}''
(\cite{cande-sciama}, p.~1717). They view the limit of large $X$ as a
means to eliminate the transient effects due to the detector starting
its inertial motion at $\tau=0$.

In our view these transient effects are already contained in the
Minkowski vacuum part (\ref{inert-Mink}) of the response, and the
additional Rindler vacuum contribution (\ref{log-div}) shows that the
inertial detector in the Rindler vacuum responds genuinely differently
than in the Minkowski vacuum. In particular, the additional Rindler
vacuum contribution (\ref{log-div}) diverges as the trajectory
approaches the Rindler horizon. This divergence was found previously
by Davies and Ottewill~\cite{davies}, both by a numerical evaluation
of the response and by a comparison of the response with the
expectation value of~$\phi^2$. The divergence can indeed be expected
on the grounds that in the Rindler vacuum the expectation values of
both $\phi^2$ and the stress-energy tensor diverge on the
horizon~\cite{candelas,davies,takagi}.

\section{Detector at rest in Newtonian gravitational field}
\label{sec:newtonian}

In this section we shall find the response of an Unruh-DeWitt detector
at rest in a static, Newtonian gravitational field: a~static,
asymptotically flat spacetime that satisfies the linearised Einstein
equations with pressureless matter as source. The quantum field is
assumed massless, but with arbitrary curvature coupling, and its
vacuum state is taken to be the Boulware-like vacuum defined in terms
of the global timelike Killing vector. We shall see that in this
situation the detector's excitation rate is always zero, but the
de-excitation rate (response for negative~$\omega$) in general differs
from that of an inertial detector in the Minkowski vacuum in Minkowski
space, and the gravitational correction depends on the details of the
mass distribution even when the detector is far from the source. We
also provide an order-of-magnitude estimate for this gravitational
correction in atomic physics decay rates on the Earth's surface.

\subsection{General matter distribution} 

The metric takes the form $g_{\mu\nu}=\eta_{\mu\nu}+h_{\mu\nu}$, where
$\eta_{\mu\nu}$ is the Minkowski metric and $h_{\mu\nu}$ is the
linearised correction. We use a system of Minkowski coordinates $(t,
\mathbf{x})$ in which $\eta_{\mu\nu} \, \mathrm{d}\mathsf{x}^\mu \,
\mathrm{d} \mathsf{x}^\nu = - \mathrm{d}t^2 + \mathrm{d}\mathbf{x}^2$
and the correction components $h_{\mu\nu}$ are independent of~$t$. We
assume $h_{\mu\nu}$ to be small enough for validity of linearised
Einstein's equations, and at large $r := \vert \mathbf{x} \vert$ we
assume $h_{\mu\nu}$ to have the asymptotically flat
falloff~$O\bigl(r^{-1}\bigr)$. We further take $h_{\mu\nu}$ to be in
the Lorentz gauge, $\partial_\mu {\bar h}^{\mu\nu}=0$, where ${\bar
h}_{\mu\nu} := h_{\mu\nu} - \frac{1}{2}
\eta_{\mu\nu} h^\alpha{}_\alpha$. 

We need the Wightman distribution in this spacetime, in the
Boulware-like vacuum that reduces to the Minkowski vacuum at the
asymptotically flat infinity. Working in perturbation theory to the
order that is consistent with linearised Einstein's equations, this
Wightman distribution must be the sum of the Minkowski vacuum
contribution ${[4\pi^2{(\Delta\mathsf{x})}^2]}^{-1}$ and a correction
$W^{(1)}(\mathsf{x},\mathsf{x}')$ that is first-order in $h_{\mu\nu}$
and dies off at infinity. Restricting the attention to a detector that
remains at constant $\mathbf{x}$ and was switched on in the infinite
past, equation (\ref{diff-rate}) shows that the transition rate reads
\begin{equation}
\label{rate-New}
\dot{F}(\omega)
=
-\frac{\omega}{2\pi}\Theta(-\omega)
+2\, \mathrm{Re}
\int_0^{\infty}\mathrm{d}s\,
\mathrm{e}^{-i\omega s}\,
W^{(1)}(t,\mathbf{x};t-s,\mathbf{x})
\,. 
\end{equation}
The first-order Wightman distribution $W^{(1)}$ in (\ref{rate-New})
depends on $\mathbf{x}$ and $s$ but not on~$t$, and from 
(\ref{eq:v-small-expansion}) we see that it has at $s=0$ 
an integrable logarithmic singularity 
proportional to the Ricci scalar. 


To find $W^{(1)}(\mathsf{x},\mathsf{x}')$, we first calculate the 
Feynman Green's function 
$G_F(\mathsf{x},\mathsf{x}')$
to first order in $h_{\mu\nu}$, adapting the procedure 
that was introduced in \cite{star} 
in the context of vacuum polarisation. 
We then find $W^{(1)}(\mathsf{x},\mathsf{x}')$ from the relation 
\begin{equation}
\label{eq:GF-vs-W}
iG_F(\mathsf{x},\mathsf{x}')
=
W(\mathsf{x},\mathsf{x}')\,\Theta(t-t')
+
W(\mathsf{x}',\mathsf{x})\,\Theta(t'-t) 
\,, 
\end{equation}
which is reliable order by order in perturbation theory as long as no
new singularities turn up, and this will be seen to be the case for
the first-order contributions.

Consider the equation satisfied by $G_F(\mathsf{x},\mathsf{x}')$, 
\begin{equation}
\left[\square_{\mathsf{x}} - \xi R(\mathsf{x})\right]
G_{F} (\mathsf{x},\mathsf{x}')
=
\frac{1}{\sqrt{-g(\mathsf{x})}}\,\delta (\mathsf{x},\mathsf{x}')
\,,
\end{equation}
and expand $g_{\mu\nu}=\eta_{\mu\nu}+h_{\mu\nu}$ and 
$G_F(\mathsf{x},\mathsf{x}')
=G_F^{(0)}(\mathsf{x},\mathsf{x}')
+G_F^{(1)}(\mathsf{x},\mathsf{x}')$, 
where $G_F^{(0)}$ is the Minkowski vacuum Feynman propagator, 
\begin{align}
G_F^{(0)}(\mathsf{x},\mathsf{x}')
= 
\frac{-i}{4\pi^2
\left[
-{(t-t')}^2
+{\vert \mathbf{x}-\mathbf{x}'\vert}^2
+i\epsilon
\right]}
\,,
\end{align}
with its distributional part 
specified by the prescription $\epsilon \to0_+$.
Dropping second order terms and noting that 
$G_F^{(0)}$ satisfies the zeroth-order equation, 
we obtain
\begin{align}
\square_{\mathsf{x}}^{(0)}\,G_F^{(1)}(\mathsf{x},\mathsf{x}')
&=
\left[ 
\partial_{\mu}\left( \bar{h}^{\mu\nu}\,\partial_\nu\right) 
+\xi \,R^{(1)}(\mathsf{x})
\right] 
G_F^{(0)}(\mathsf{x},\mathsf{x}')
\nonumber
\\
&= 
\left[ \, 
{\bar{h}}^{00}\partial_t^2
+\xi \,R^{(1)}(\mathsf{x})
\right] 
G_F^{(0)}(\mathsf{x},\mathsf{x}')
\,, 
\label{eqGF}
\end{align}
where the last equality follows because 
in the Lorentz gauge ${\bar h}_{\mu\nu}$ satisfies the Minkowski space 
wave equation
\begin{align}
\label{eq:lin-einstein}
\square^{(0)}
{\bar h}_{\mu\nu} = - 16 \pi G T_{\mu\nu}
\end{align}
and we are assuming static, pressureless matter. 

The key observation now is that because the spacetime is static and
asymptotically flat, $G_F$ can be regarded as the analytic
continuation of the \emph{unique\/} Green's function on the positive
definite section. As this holds order by order in perturbation theory,
we can solve (\ref{eqGF}) for $G_F^{(1)}$ by using $G_F^{(0)}$ as the
inverse of~$\square^{(0)}$, with the result
\begin{align}
G_F^{(1)} (\mathsf{x},\mathsf{x}')
&=
\int \mathrm{d}
\tilde{\mathsf{x}}\,
G_F^{(0)}(\mathsf{x},\tilde{\mathsf{x}})
\left[ 
\bar{h}^{00}(\tilde{\mathsf{x}})\partial_{\tilde{t}}^2
+\xi \,R^{(1)}(\tilde{\mathsf{x}})
\right] 
G_F^{(0)}(\tilde{\mathsf{x}},\mathsf{x}')
\nonumber
\\
&=
-\frac{1}{16\pi^4}
\int \mathrm{d}
\tilde{\mathsf{x}} \, 
\frac{1}{
\left[
-{(t-\tilde{t}\,)}^2
+{\vert \mathbf{x}-\tilde{\mathbf{x}}\vert}^2
+i\epsilon 
\right]
} 
\nonumber
\\[1ex] 
&\hspace{15ex}
\times
\left[ 
\bar{h}^{00}(\tilde{\mathsf{x}})
\partial_{\tilde{t}}^2+\xi \,R^{(1)}(\tilde{\mathsf{x}})
\right]
\frac{1}
{
\left[
-{(t'-\tilde{t}\,)}^2
+{\vert \mathbf{x}'-\tilde{\mathbf{x}}\vert}^2+i\epsilon
\right]}
\,. 
\end{align}
The integral over 
$\tilde{t}$ can be done by residues, 
using the time-independence of $R^{(1)}$ and~$\bar{h}^{00}$. 
Defining 
$G_F^{(1)} (s,\mathbf{x}) := G_F^{(1)}(t,\mathbf{x};t-s,\mathbf{x})$, 
and writing $X := \vert \mathbf{x}-\tilde{\mathbf{x}}\vert$ for short, 
we obtain 
\begin{align}
G_F^{(1)} (s,\mathbf{x})
&=
\frac{-i}{8\pi^3}
\int
\frac{\mathrm{d}
\tilde{\mathbf{x}}}
{
\sqrt{X^2 + i \epsilon}}
\, 
\Biggl[
\frac{\xi\,R^{(1)}(\tilde{\mathbf{x}})}{
s^2- 4 \left( X^2 
+ i \epsilon \right)  
}
+\frac{2\bar{h}^{00}(\tilde{\mathbf{x}}) 
\left[
3s^2 + 4 \left( X^2 
+ i \epsilon \right) \right]
}
{{
\left[
s^2- 4 \left( X^2 
+ i \epsilon \right) 
\right]
}^3 }
\Biggr]
\,.
\end{align}
As $\nabla^2 \bar{h}^{00}=-2R^{(1)}$, where 
$\nabla^2 := 
\partial^2_{x_1}
+ 
\partial^2_{x_2} 
+ 
\partial^2_{x_3}$, 
and as $h_{\mu\nu}$ has the falloff~$O\bigl(r^{-1}\bigr)$, we can
integrate the term involving $\bar{h}^{00}$ by parts and take
$R^{(1)}$ as a common factor. Assuming $s\ne0$, and dropping terms
that go to zero as $\epsilon \to0$, we obtain
\begin{align}
& G_F^{(1)} (s,\mathbf{x})
=
\frac{-i}{8\pi^3}\int
\mathrm{d}
\tilde{\mathbf{x}}\,
R^{(1)} (\tilde{\mathbf{x}})
\Bigg[
\frac{2\xi -1}{2X
\left[
\left( s^2- 4i\epsilon \right)
- 4 X^2 
\right]}
+
\frac{1}
{X {\left( s^2- 4i\epsilon \right)} }
\nonumber
\\[1ex]
&
\hspace{35ex}
+ 
\frac{1}{{\left(s^2- 4i\epsilon\right)}^{3/2}}
\ln\left(\frac{2X
- \sqrt{s^2- 4i\epsilon}}
{2X
+ \sqrt{s^2- 4i\epsilon}} \right)
\Bigg]
\,. 
\label{GF}
\end{align}
Note that the $O\bigl(r^{-1}\bigr)$ falloff of $h_{\mu\nu}$ 
implies that $R^{(1)}$ has the falloff~$O\bigl(r^{-3}\bigl)$, 
and the integral in (\ref{GF}) hence converges pointwise in~$s$. 
As a consistency check, we also note that 
if $R^{(1)}$ vanishes in a neighbourhood of the point~$\mathbf{x}$, 
taking in (\ref{GF}) the limit $s\to0$ and 
$\epsilon\to0$ 
yields a nonsingular expression that 
reproduces the vacuum polarisation 
$\langle \phi^2(\mathbf{x})\rangle$ 
that was found 
for this class of spacetimes in~\cite{star}. 

We now assume $s>0$. Let $G_{F,1}^{(1)} (s,\mathbf{x})$ and
$G_{F,2}^{(1)} (s,\mathbf{x})$ denote the contributions to
$G_F^{(1)} (s,\mathbf{x})$ from respectively the first term and the
last two terms in the integral in~(\ref{GF}). Taking the limit
$\epsilon\to0$ in $G_{F,2}^{(1)} (s,\mathbf{x})$ is
elementary, with the result
\begin{align}
G_{F,2}^{(1)} (s,\mathbf{x})
& =
- 
\frac{i}{2\pi^2} \int_0^\infty 
\mathrm{d} X \, 
{\tilde R}(X) 
\left\{
\frac{X}{s^2}
+ 
\frac{X^2}{s^3}
\left[ 
\ln\left(\frac{ | 2X
- s |}
{2X + s} \right)
+ i \pi \Theta \! \left(\frac{s}{2X} -1 \right)
\right]
\right\}
\,, 
\end{align} 
where ${\tilde R}(X)$ denotes the average of 
$R^{(1)}$ over a sphere of radius $X$ about $\mathbf{x}$ 
and the dependence of ${\tilde R}(X)$ on $\mathbf{x}$ is suppressed. 
In $G_{F,1}^{(1)} (s,\mathbf{x})$, 
splitting the integrand into partial fractions and taking the limit 
$\epsilon\to0$ yields 
\begin{align}
\label{eq:G1-int1}
G_{F,1}^{(1)} (s,\mathbf{x})
& =
\frac{2\xi -1}{32\pi} 
\left[
{\tilde R}(s/2) 
+ i ( H {\tilde R} ) (s/2) 
+ i ( H {\tilde R} ) (-s/2) 
\right]
\,, 
\end{align} 
where $H$ stands for the Hilbert transform, 
\begin{align}
(Hf) (x) := 
\frac{1}{\pi} \, P \! \int_{-\infty}^\infty 
\mathrm{d}y
\, 
\frac{f(y)}{y-x}
\, , 
\end{align}
with $P$ denoting the principal value integral, and ${\tilde R}$ is
understood to vanish for negative argument. Recall now that the
Hilbert transform can be written as $Hf = -i f_+ + i f_-$, where 
$f_+$ and $f_-$ are respectively the projections of $f$ to the
positive and negative frequency subspaces, 
$f_+(x) := {(2\pi)}^{-1/2}
\int_0^\infty {\mathrm{e}}^{-i\omega x} 
{\hat f}(\omega) \, \mathrm{d}\omega$, 
$f_-(x) := {(2\pi)}^{-1/2}
\int_{-\infty}^0 {\mathrm{e}}^{-i \omega x} {\hat f}(\omega) \,
\mathrm{d}\omega$, and ${\hat f}$ denotes the Fourier transform, 
${\hat f}(\omega) = {(2\pi)}^{-1/2}
\int_{-\infty}^\infty {\mathrm{e}}^{i \omega x} f(x) \,
\mathrm{d}x$ \cite{titchmarsh-fourier}. 
As ${\tilde R}$ vanishes for negative argument, it thus follows from
(\ref{eq:G1-int1}) that
\begin{align}
\label{eq:G1-int2}
G_{F,1}^{(1)} (s,\mathbf{x})
& =
\frac{2\xi -1}{16\pi} 
\left[
{\tilde R}_+(s/2) 
+ {\tilde R}_+ (-s/2) 
\right]
\,. 
\end{align}

\begin{figure}[t]
\centering
\includegraphics[width=0.8\textwidth]{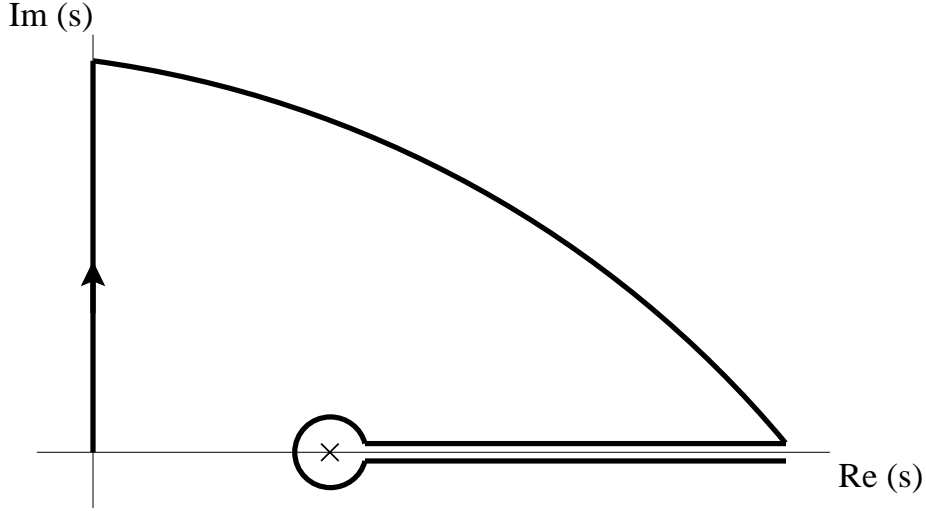}
\caption{The contour in the complex $s$ plane for evaluating the 
$s$-integral in (\ref{eq:F2sss}) for $\omega<0$. 
The branch point at $s = 2X$ is indicated by a cross 
and the cut is at $s>2X$.}
\label{fig:contour}
\end{figure}

Let now $\Delta \dot{F}_{\mathbf{x}}(\omega) := 
\dot{F}(\omega) + 
\bigl[(\omega/(2\pi)\bigr]\Theta(-\omega)$ denote 
the correction to the Minkowski space transition rate. 
Using (\ref{rate-New}) and (\ref{eq:GF-vs-W}), we can write 
$\Delta \dot{F}_{\mathbf{x}}(\omega)$ in terms of 
$G_{F,1}^{(1)}$ and $G_{F,2}^{(1)}$ as 
\begin{equation}
\Delta \dot{F}_{\mathbf{x}}(\omega)
=
2\, \mathrm{Re}
\int_0^{\infty}\mathrm{d}s\,
\mathrm{e}^{-i\omega s}\,
\left[
i G_{F,1}^{(1)} (s,\mathbf{x})
+ 
i G_{F,2}^{(1)} (s,\mathbf{x})
\right] 
\,. 
\end{equation}
The contribution to 
$\Delta \dot{F}_{\mathbf{x}}(\omega)$ from 
$G_{F,1}^{(1)}$ equals 
\begin{align}
\Delta \dot{F}_{\mathbf{x},1}(\omega)
& = 
- \frac{2\xi -1}{8\pi} 
\, \mathrm{Im} \,  
\int_0^{\infty}\mathrm{d}s\,
\mathrm{e}^{-i\omega s}\,
\left[ 
{\tilde R}_+(s/2) 
+
{\tilde R}_+(-s/2) 
\right]
\nonumber 
\\
& = 
- \frac{2\xi -1}{8\pi} 
\, \mathrm{Im} \,  
\int_{-\infty}^{\infty}\mathrm{d}s\,
\mathrm{e}^{-i\omega s}\,
{\tilde R}_+(s/2) 
\nonumber 
\\
& = 
- \frac{2\xi -1}{8\pi} 
\, 
\Theta(-\omega) \, 
\mathrm{Im} \, 
\int_{-\infty}^{\infty}\mathrm{d}s\,
\mathrm{e}^{-i\omega s}\,
{\tilde R}(s/2) 
\nonumber 
\\
& = 
- \frac{2\xi -1}{4\pi} 
\, 
\Theta(-\omega) \, 
\mathrm{Im} \, 
\int_0^{\infty}\mathrm{d}X\,
\mathrm{e}^{-2i\omega X}\,
{\tilde R}(X) 
\nonumber 
\\
& = 
\frac{2\xi -1}{16\pi^2} 
\, 
\Theta(-\omega) \, 
\int \mathrm{d} \tilde{\mathbf{x}}
\, 
R^{(1)} (\tilde{\mathbf{x}})
\, 
\frac{\sin(2\omega X)}{X^2}
\,, 
\label{eq:F1}
\end{align}
where we have used the definition of the positive frequency
projection, changed the integration variable to $X = s/2$ and finally
written ${\tilde R}$ in terms of~$R^{(1)}$. To evaluate the
contribution to $\Delta \dot{F}_{\mathbf{x}}(\omega)$ from
$G_{F,2}^{(1)}$, we interchange the integrals over $s$ and $X$,
justified by the absolute convergence of the double integral, and
obtain
\begin{align}
\Delta \dot{F}_{\mathbf{x},2}(\omega)
& = 
\frac{1}{\pi^2} 
\int_0^\infty 
\mathrm{d} X \, 
{\tilde R}(X) 
\nonumber
\\
& \hspace{3ex}
\times 
\mathrm{Re}
\int_0^{\infty}\mathrm{d}s\,
\mathrm{e}^{-i\omega s}\,
\left\{
\frac{X}{s^2}
+ 
\frac{X^2}{s^3}
\left[ 
\ln\left(\frac{ | 2X -s |}
{2X+ s} \right)
+ i \pi \Theta \! \left(\frac{s}{2X} -1 \right)
\right]
\right\}
\,. 
\label{eq:F2sss}
\end{align}
The integral over $s$ in (\ref{eq:F2sss}) may be interpreted as the
integral of a complex analytic function along the positive real axis,
with a contour deformation to the lower half-plane near the
logarithmic singularity at $s = 2X$. For $\omega>0$, the contour can
be deformed to the negative imaginary axis, and the integral vanishes
on taking the real part. For $\omega<0$, the contour can be deformed
to that shown in Figure~\ref{fig:contour}. The contribution from the
large arc vanishes when the arc is taken to infinity, and the
contribution from the positive imaginary axis vanishes on taking the
real part. The only nonvanishing contribution comes from the branch
cut at $s>2X$. Collecting, we find
\begin{align}
\Delta \dot{F}_{\mathbf{x},2}(\omega)
& = 
\frac{2}{\pi}
\, \Theta(-\omega) 
\int_0^\infty 
\mathrm{d} X \, X^2 \, 
{\tilde R}(X) 
\int_{2X}^{\infty}\mathrm{d}s\,
\frac{\sin(\omega s)}{s^3}
\nonumber
\\[1ex]
& = 
\frac{1}{2\pi^2}
\, \Theta(-\omega) 
\int \mathrm{d} \tilde{\mathbf{x}}
\, 
R^{(1)} (\tilde{\mathbf{x}})
\int_{2X}^{\infty}\mathrm{d}s\,
\frac{\sin(\omega s)}{s^3}
\,. 
\label{eq:F2}
\end{align}
If desired, the integral over $s$ in (\ref{eq:F2}) can be expressed as
a sum of elementary functions and the sine integral function. The form
in (\ref{eq:F2}) is however more convenient for the observations that
we shall make below.

Combining (\ref{eq:F1}) and~(\ref{eq:F2}), and using the linearised
Einstein equation (\ref{eq:lin-einstein}) to write $R^{(1)} = 8\pi G
\rho$, where $\rho$ is the matter density, the gravitational
correction to the Minkowski space transition rate takes the final form
\begin{equation}
\label{rate-gen}
\Delta \dot{F}_{\mathbf{x}}(\omega)
=
\frac{G}{2\pi}\,\Theta(-\omega)
\int\mathrm{d}
\tilde{\mathbf{x}}\,
\rho(\tilde{\mathbf{x}})
\left[
(2\xi -1) \,\frac{\sin(2\omega X)}{X^2}
+ 8 \int_{2X}^{\infty}\mathrm{d}s\,
\frac{\sin(\omega s)}{s^3}
\right]
\,, 
\end{equation}
where we recall that $X$ is defined by $X:=\vert
\mathbf{x}-\tilde{\mathbf{x}}\vert$. As a check, we note that the
integral in (\ref{rate-gen}) converges in absolute value: the quantity
in the brackets is is of order $O\bigl(X^{-2}\bigr)$ as $X\to\infty$
and of order $O\bigl(X^{-1}\bigr)$ as $X\to0$, and $\rho(\mathbf{x}) =
O\bigl({\vert \mathbf{x}\vert}^{-3}\bigr)$ at $\vert \mathbf{x}\vert
\to\infty$.

Two observations on the gravitational 
correction (\ref{rate-gen}) are immediate. 
First, the correction vanishes for $\omega>0$. 
The excitation rate is thus zero: 
\emph{a~static Newtonian gravitational field 
causes no excitations in a static detector.} 

Second, the correction depends on the matter distribution within the
source even when the source is compact and we consider the leading
behaviour far from the source,
$\vert\mathbf{x}\vert\rightarrow\infty$. Classical intuition might
suggest that in this limit the de-excitation rate should only depend
on the monopole moment of the mass distribution, the total
mass. However, as the second term in the brackets in (\ref{rate-gen})
is of order $O\bigl(X^{-3}\bigr)$ at $X\to\infty$, the leading
contribution at $\vert\mathbf{x}\vert\rightarrow\infty$ comes entirely
from the first term in the brackets and can be evaluated, with the
result
\begin{align}
\Delta \dot{F}_{\mathbf{x}}(\omega)
& 
=
\frac{G(2\xi -1)}{2\pi {\vert\mathbf{x}\vert}^2}\,\Theta(-\omega)
\biggl[
\sin\bigl(2\omega \vert\mathbf{x}\vert\bigr)
\int\mathrm{d}
\tilde{\mathbf{x}}\,
\rho(\tilde{\mathbf{x}})
\cos(2\omega \, \hat{\mathbf{x}} \cdot \tilde{\mathbf{x}})
\nonumber\\
&
\hspace{10ex}
- 
\cos\bigl(2\omega \vert\mathbf{x}\vert\bigr)
\int\mathrm{d}
\tilde{\mathbf{x}}\,
\rho(\tilde{\mathbf{x}})
\sin(2\omega \, \hat{\mathbf{x}} \cdot \tilde{\mathbf{x}})
\biggr]
+ O \bigl( {\vert\mathbf{x}\vert}^{-3} \bigr)
\,, 
\label{rate-asympt}
\end{align}
where $\hat{\mathbf{x}} := \mathbf{x}/\vert \mathbf{x} \vert$. If
$\vert \omega R_0\vert\ll 1$, where $R_0$ is the characteristic length
scale of the source, the square brackets in (\ref{rate-asympt}) can be
approximated by $M \sin\bigl(2\omega \vert\mathbf{x}\vert\bigr)$,
where $M$ is the total mass, but outside this limit the square
brackets depend also on the higher multipole moments of the source. It
is worth mentioning that when $R_0$ is a typical stellar or planetary
scale and $\omega$ is a typical atomic frequency, we in fact have
$\vert \omega R_0\vert \gg1$, in which limit the integrals in
(\ref{rate-asympt}) can be estimated by WKB techniques. The
de-excitation rate of an atom far from a star therefore carries an
imprint of the internal structure of the star.

\subsection{Constant density star} 
\label{subsec:constdensity} 

As an example, we consider 
a spherical star of constant density 
$\rho_0$ and radius~$R_0$. 
The Ricci scalar has now a discontinuity at the surface of the star, 
and this spacetime therefore 
falls outside the smooth setting in which we have been working. 
We suspect that the non-smoothness is a technical issue that 
will not have a significant effect on 
the particle detector and shall proceed, 
with due caution. 

The integrals in 
(\ref{rate-gen}) can be evaluated in terms of the sine integral 
function. 
When $\mathbf{x}$ is outside the star, 
$r := \vert\mathbf{x}\vert > R_0$, we find 
\begin{align}
\label{star-result}
&\Delta\dot{F}_{r}(\omega)=\frac{G\rho_0}{4\omega^2 r}\Theta(-\omega)
\Bigg[\xi\Bigg(
2\omega(r+R_0)\cos[2\omega(r-R_0)]-2\omega(r-R_0)\cos[2\omega(r+R_0)]\nonumber\\
&+\sin[2\omega(r-R_0)]-\sin[2\omega(r+R_0)]
\nonumber\\
&+4\omega^2(r^2-R_0^2)\Big( \mathrm{Si}[2\omega(r-R_0)]-\mathrm{Si}[2\omega(r-R_0)]\Big) 
\Bigg)\nonumber\\
&+\frac{1}{6}\Bigg(
64\,\pi\omega^4 r R_0^3-\Big(2\omega (r+3 R_0)-4\omega^3(r^3+r^2 R_0-5r R_0^2+3R_0^3)\Big)\cos[2\omega(r-R_0)]\nonumber\\
&+\Big(2\omega (r-3 R_0)+4\omega^3(-r^3+r^2 R_0+5r R_0^2+3R_0^3)\Big)\cos[2\omega(r+R_0)]\nonumber\\
&+\Big(-3+2\omega^2(r^2+2r R_0-3R_0^2)\Big)\sin[2\omega(r-R_0)]
\nonumber\\
&+\Big(3+2\omega^2(-r^2+2r R_0-3R_0^2)\Big)
\sin[2\omega(r+R_0)]
\nonumber\\
&+8\omega^4(r-R_0)^3(r+3R_0)\mathrm{Si}[2\omega(r-R_0)]
-8\omega^4(r+R_0)^3(r-3R_0)\mathrm{Si}[2\omega(r+R_0)]
\Bigg)
\Bigg]
\,. 
\end{align}
Consider in particular the limit in which $\mathbf{x}$ approaches the
surface of the star, $r\to R_0$: this presumably is the situation with
the best experimental prospects of observing the gravitational
correction to the de-excitation rate. We find
\begin{align}
\Delta\dot{F}_{R_0}(\omega)
&=
\frac{G\rho_0}{48\omega^2 R_0}\Theta(-\omega)
\Big[
4\omega R_0(8\omega^2R_0^2-1)\cos(4\omega R_0)
+(3-12\xi+8\omega^2R_0^2)\sin(4\omega R_0)
\nonumber\\
&
\hspace{5ex}
+8\omega R_0(8\pi\omega^3 R_0^3+6\xi-1)
+128\omega^4R_0^4\,\mathrm{Si}(2\omega R_0)
\Big]
\,. 
\label{eq:rate-surface}
\end{align}
For $\vert \omega R_0\vert\gg 1$, the asymptotic behaviour  
of (\ref{eq:rate-surface}) is 
\begin{equation}
\label{eq:rate-surface-largefreq}
\Delta\dot{F}_{R_0}(\omega)
\sim
\frac{G\rho_0}{\omega}\left( \xi-\frac{1}{6}\right) 
\Theta(-\omega)
\,. 
\end{equation}
If $\rho_0$ is the density of the Earth and 
$\omega$ is a typical atomic frequency, the ratio of 
(\ref{eq:rate-surface-largefreq}) to the Minkowski vacuum 
transition rate $[\omega/(2\pi)]\Theta(-\omega)$
is of order $(G\rho_0/\omega^2)
\sim 
10^{-42}$. 
If the transition rate for the electromagnetic field 
behaves qualitatively similarly to that in our scalar field model, 
we conclude that the gravitational correction 
to decay rates in atomic physics laboratory 
experiments is unobservably small.

\section{Conclusions}
\label{sec:conclusions} 

In this paper we have discussed the instantaneous transition rate of
an Unruh-DeWitt detector that is coupled to a scalar field in an
arbitrary Hadamard state in curved spacetime. We started with a
detector that is switched on and off smoothly, in which case the
detector's response function $F(\omega)$ is well defined and can be
expressed as the integral of a Wightman distribution with a standard
$i\epsilon$ regulator. We showed that the limit $\epsilon\to0$ can be
taken explicitly, with the result
\begin{align}
F(\omega) 
&=
-\frac{\omega}{4\pi}\int_{-\infty}^{\infty}\mathrm{d}u\,{[\chi(u)]}^2 
\ + \ 
\frac{1}{2\pi^2}\int_0^{\infty}
\frac{\mathrm{d}s}{s^2}\int_{-\infty}^{\infty}\mathrm{d}u\,\chi(u)
\bigl[ \chi(u)-\chi(u-s)\bigr] 
\nonumber 
\\
\noalign{\medskip}
&\hspace{3ex}
+ 
2
\int_{-\infty}^{\infty}\mathrm{d}u\,\chi(u)
\int_0^{\infty}\mathrm{d}s\,\chi(u-s) 
\,\mathrm{Re} \left[ 
\mathrm{e}^{-i\omega s}\, W_0(u,u-s)+\frac{1}{4\pi^2s^2}
\right] 
\,, 
\label{probability-conc}
\end{align}
where $\chi$ is the switching function and $W_0$ is the pull-back of
the Wightman distribution to the detector's world line, with the
$\epsilon$-regulator having been taken pointwise to zero. We then
showed that when the switch-on and switch-off have a fixed shape but
take each place within the time interval~$\delta$, the sharp switching
limit $\delta\to0$ results into a logarithmic divergence in
$F(\omega)$, but the derivative of $F(\omega)$ with respect to the
total detection time $\Delta\tau$ remains finite and is given by
\begin{equation}
\label{resultado1-conc}
\dot{F}_{\tau}(\omega)
=
-\frac{\omega}{4\pi}
+2\int_0^{\Delta\tau}\mathrm{d}s
\, \mathrm{Re}
\left[ \mathrm{e}^{-i\omega s}W_0(\tau,\tau-s)
+\frac{1}{4\pi^2s^2}\right]
\ \ +\frac{1}{2\pi^2 \Delta \tau} + O(\delta)  
\, . 
\end{equation}
As a consequence, the difference $\Delta \dot{F}_{\tau}(\omega)$ in
the transition rates of two detectors in different quantum states, on
different trajectories and even in different spacetimes, but having
the same switching function, has a well defined $\delta\to0$ limit,
given by
\begin{equation}
\label{diff-rate-conc}
\Delta \dot{F}_{\tau}(\omega) 
= 
2\, \mathrm{Re}
\int_0^{\Delta\tau}\mathrm{d}s\,\mathrm{e}^{-i\omega s}
\left[
W_0^A(\tau,\tau-s)-W_0^B(\tau,\tau-s)
\right]\,. 
\end{equation}
The case of a detector switched on in the infinite past can be 
defined by the  
$\Delta\tau\rightarrow\infty$ limit in 
(\ref{resultado1-conc}) and~(\ref{diff-rate-conc}), 
subject to suitable asymptotic conditions. 
We emphasise that all the integrals in the above formulas 
are integrals of ordinary functions, no longer involving $i\epsilon$
regulators or other distributional aspects. These results generalise
to the setting of general Hadamard states in curved spacetime the
results that were obtained for the massless field in the Minkowski
vacuum in~\cite{switching}.

We applied the difference formula (\ref{diff-rate-conc}) to two
situations in which the reference state can be conveniently chosen to
be an inertial detector in the Minkowski vacuum. First, we considered
an inertial detector coupled to a massless field in the Rindler vacuum
in Minkowski space, finding that the transition rate diverges
logarithmically as the detector approaches the Rindler
horizon. Second, we considered a detector at rest in a static,
Newtonian gravitational field, coupled to a massless field with
arbitrary curvature coupling, in the Boulware-like vacuum defined with
respect to the global timelike Killing vector. We found the excitation
rate to be zero, but the de-excitation rate acquires a gravitational
correction that depends on the details of the mass distribution within
the source, even in the limit in which the source is compact and the
detector is far from the source. Using a spherical constant density
mass distribution as an example, we estimated the gravitational
corrections to decay rates in atomic physics laboratory experiments on
the surface of the Earth to be suppressed by 42 orders of magnitude.

A~technical assumption throughout the paper was that both the
spacetime and the detector trajectory were taken smooth. Given that
the final formulas (\ref{probability-conc}) and
(\ref{resultado1-conc}) remain well defined whenever the trajectory is
sufficiently differentiable for the $s^{-2}$ term to subtract the
non-integrable part in~$W_0$, it is tempting to suspect that the
smoothness assumption on the trajectory could be relaxed. To
investigate this question, two steps would need to be addressed.
First, in section \ref{sec:detectors} we justified the use of the
$i\epsilon$-regulator in the pull-back of the Wightman distribution in
(\ref{defresponse-eps}) by the theorems
of~\cite{hormander-vol1,hormander-paper1}, which are formulated for
smooth submanifolds: how do these theorems generalise to a lower
degree of differentiability?\footnote{This issue arises already in the
Minkowski vacuum analysis in~\cite{switching}.} Second, to obtain in
section \ref{sec:eps-limit} the estimates in~(\ref{expansion}), we
assumed a trajectory that is $C^8$ and has a suitably bounded
remainder term in the Taylor expansion (say, $C^9$ would suffice):
could the techniques of section \ref{sec:eps-limit} be improved to
relax this assumption?

Our results provide tools for investigating a particle detector's
response in time-dependent situations in curved spacetime. One~set of
questions with which these tools could prove useful are thermal
effects in black hole spacetimes in the time-dependent setting. For
example, a detector that is falling freely into a static black hole in
a Boulware-type vacuum \cite{bo:boulware} would be expected to have a
divergent response at the horizon, in analogy with the Rindler vacuum
analysis in our section~\ref{sec:rindler}, but might any thermal
characteristics survive in the response of a detector falling through
the horizon in an Unruh-type state \cite{unruh} or in a
Hartle-Hawking-Israel type state
\cite{hawking,israel-vacuum}? 
From a complementary angle, consider the spacetime of a collapsing
star, in a quantum state that was Boulware-type in the distant past:
how does the response of a detector at a fixed position outside the
star evolve from that found in section \ref{sec:newtonian} to the
thermal response? 
In particular, what is the time scale of this evolution, and 
how is the detector's response in this situation related to the 
the outoing energy flux that develops, or to any notion of 
`particles' in the associated Bogoliubov transformation? 
On a more speculative note, might there be a relation 
between the response of a detector and 
dynamical or evolving horizons~\cite{ashtekar-dynamical}?

\section*{Acknowledgements}
We thank Chris Fewster for extremely helpful discussions, 
and in
particular for making us aware of the results in  
\cite{hormander-vol1,hormander-paper1}. 
AS thanks Bill Unruh for helpful suggestions and hospitality
at the conception stages of this work.
This work was supported in part by 
STFC (UK) Rolling Grant PP/D507358/1. 
AS was supported by an EPSRC Dorothy Hodgkin 
Research Award to the University of Nottingham.

\begin{appendix}

\section{Appendix: 
Divergence of the transition rate at the Rindler horizon}

In this appendix we verify the 
logarithmically divergent asymptotic form 
(\ref{explicit-logdiv-text})
for the difference 
$\Delta \dot{F}_{\tau}(\omega)$ 
of the inertial detector transition rates in the Rindler vacuum and
the Minkowski vacuum as the detector approaches the Rindler horizon.

We start from formula (\ref{log-div}) 
for $\Delta \dot{F}_{\tau}(\omega)$. 
Writing  
$\tau = X(1-\epsilon)$ and scaling the integration variable in 
(\ref{log-div}) by $s\rightarrow Xs$, we have 
\begin{align}
\label{log-div-scaled}
\Delta \dot{F}_{\tau}(\omega) 
&=
\frac{1}{2\pi^2 X}
\int_0^{\alpha - \epsilon}\mathrm{d}s\,
\frac{\cos(\beta s)}{s^2}
\left[
1+\frac{s}{2(1-\epsilon)-s}
\left( \frac{1}{\ln\left(1+\frac{s}{\epsilon}\right)} 
+\frac{1}{\ln\left( 1-\frac{s}{2-\epsilon}\right) } 
\right)  
\right] 
\,,
\end{align}
where $\beta:= \omega X$ 
and $\alpha := 1 - (\tau_0/X)$. We shall find the 
asymptotic form of (\ref{log-div-scaled}) for 
$\epsilon\to0_+$ with fixed $X$, $\alpha$ and~$\beta$. 
Note that $0<\alpha <2$. 

We note first that we may replace the upper limit of the integral in
(\ref{log-div-scaled}) by $\alpha$ at the expense of an error of
order~$O(\epsilon)$. To handle the remaining integral, we split the
interval $(0,\alpha)$ into the subintervals $(0,\eta)$ and
$(\eta,\alpha)$, where $\eta := \sqrt{\epsilon}$. We denote the
contributions from the two subintervals
by respectively 
${(2\pi^2X)}^{-1} I_1$ and ${(2\pi^2X)}^{-1} I_2$ 
and provide separate estimates for each.
A key tool for controlling the logarithms will be the Laurent expansion
\begin{align}
\label{eq:invlog-laurent}
\frac{1}{\ln(1+x)} = \frac{1}{x} + \frac12 + O(x)
\,. 
\end{align}

Consider~$I_1$. 
We introduce the new integration variable $r:=s/\eta$, 
with the range $0<r<1$. We replace the 
second logarithm in the integrand by 
the first two terms in~(\ref{eq:invlog-laurent}), at the expense of 
an error of order $O(\eta)$ in~$I_1$, and find  
\begin{align}
\label{eq:jjj1}
I_1
&=
\int_0^1\frac{\cos(\beta \eta r) \, \mathrm{d}r}
{r \bigl[ 2(1-\eta^2)-\eta r \bigr]}
\left( 
\frac{1}{\ln \! \left(1+ \frac{r}{\eta}\right)}-\frac{\eta}{r}-\frac{1}{2} 
\right) 
\ \ \ +O(\eta)
\,. 
\end{align}
As the quantity in the large parentheses in (\ref{eq:jjj1}) is bounded, 
we may make the replacements $\cos(\beta \eta r) \to 1$ 
and 
${\bigl[ 2(1-\eta^2)-\eta r \bigr]}^{-1}
\to 
{\bigl[ 2(1-\eta^2)\bigr]}^{-1}$ 
at the expense of respective errors of order 
$O(\eta^2)$ and~$O(\eta)$. 
Changing the integration variable to $y := r/\eta$, we then obtain 
\begin{align}
I_1
&=
\frac{1}{2(1-\eta^2)}
\int_0^{1/\eta}
\frac{\mathrm{d}y}{y}
\left(
\frac{1}{\ln\left(1+y\right)}-\frac{1}{y}- \frac{1}{2}
\right) 
\ \ \ +O(\eta)
\,. 
\label{eq:jjj5}
\end{align}

We now concentrate on the divergent part. With errors of order $O(1)$,
we first replace the lower limit of integration in (\ref{eq:jjj5})
by~$1$, then drop the $y^{-2}$ term and make the replacement ${[y \ln
\left(1+y\right)]}^{-1}
\to 
{[(1+y) \ln \left(1+y\right)]}^{-1}$. 
The remaining integral is elementary, with the result 
\begin{equation}
\label{I1}
I_1
=
\tfrac{1}{8}\ln\epsilon+\tfrac{1}{2}\ln\left( -\ln\epsilon\right) 
+O(1)
\,.
\end{equation}

Consider then~$I_2$. In the integrand shown in~(\ref{log-div-scaled}),
we replace the argument of the second logarithm by $\left(1-
\frac{s}{2}\right)$, the fraction $s/\bigl[2(1-\epsilon)-s\bigr]$ by
$s/(2-s)$ and the argument of the first logarithm by~$s/\epsilon$,
with elementary estimates showing that each step produces in $I_2$ an
error of order~$O(\eta)$, and obtain
\begin{equation}
\label{eq:jjj6}
I_2
=
\int_{\eta}^{\alpha}
\mathrm{d}s\,
\frac{\cos(\beta s)}{s^2}
\left[
1+\frac{s}{2-s}
\left( 
\frac{1}{\ln\bigl(s/\eta^2\bigr)} 
+\frac{1}{\ln\left( 1-\frac{s}{2}\right)} 
\right)  
\right] 
\ \ \ +O(\eta)
\,.
\end{equation}
The terms in (\ref{eq:jjj6}) that do not involve
$\ln\bigl(s/\eta^2\bigr)$ can be handled by a straightforward small
$s$ Laurent expansion, and their contribution to $I_2$ is $\frac14
\ln\eta + O(1)$. In the term involving $\ln\bigl(s/\eta^2\bigr)$, the
replacements $\cos(\beta s) \to 1$ and $s/(2-s) \to s/2$ can be
verified to produce in $I_2$ errors of order
$O\bigl(1/(\ln\eta)\bigr)$, and the remaining integral is elementary
and evaluates to~$O(1)$. Hence
\begin{equation}
\label{I2}
I_2
=
\tfrac{1}{8}\ln\epsilon 
+O(1)
\,.
\end{equation}

Combining these results, we have 
\begin{equation}
\label{explicit-logdiv}
\Delta \dot{F}_{\tau}(\omega)
=
\frac{1}{2\pi^2X}
\left[ 
\tfrac{1}{4} \ln\epsilon
+
\tfrac{1}{2} \ln\left( -\ln\epsilon \right) 
+O(1)
\right] \,.
\end{equation}
Formula (\ref{explicit-logdiv-text}) follows by substituting 
$\epsilon = 1 - (\tau/X)$.

\end{appendix}

\end{document}